\documentclass[preprint,12pt,authoryear,review]{elsarticle}

\usepackage{lineno}

\usepackage{graphicx}
\usepackage{amsmath}
\usepackage{natbib}
\usepackage{amssymb}
\usepackage{epstopdf}

\begin{document}

\begin{frontmatter}

\title{Melt-Enhanced Rejuvenation of Lithospheric Mantle: Insights from the Colorado Plateau}
\author[1]{Mousumi Roy}
 
\author[1,3]{Rodrigo Osuna Orozco}
\author[2]{Ben Holtzman}
\author[2]{James Gaherty}

\address[1]{Department of Physics and Astronomy, University of
New Mexico, Albuquerque, NM 87106, USA.}

\address[2]{Lamont Doherty Earth Observatory, Columbia University, Palisades, NY 10964, USA.}

\address[3]{now at Scripps Institution of Oceanography, University of California San Diego, La Jolla, CA 92037, USA.}

\begin{abstract}
The stability of the lithospheric mantle beneath the ancient cratonic cores of continents is primarily a function of chemical modification during the process of melt extraction.  Processes by which stable continental lithosphere may be destabilized are not well-understood, although destabilization by thickening and removal of negatively-buoyant lithospheric mantle in ``delamination'' events has been proposed in a number of tectonic settings.  In this paper we explore an alternative process for destabilizing continents, namely, thermal and chemical modification during infiltration of metasomatic fluids and melts into the lithospheric column.  We consider observations pertinent to the structure and evolution of the Colorado Plateau within the western United States to argue that the physical and chemical state of the margins of the plateau have been variably modified and destabilized by interaction with melts.  In the melt-infiltration process explored here, the primary mechanism for weakening and rejuvenating the plate is through thermal effects and the feedback between deformation and melt-infiltration.  We speculate on the nature and geometry of a melt-modulated interaction zone between lithosphere and asthenosphere and the seismically-observable consequences of variable melt-infiltration into the margins of regions of thick, stable lithosphere such as the Colorado Plateau and the Archean Wyoming Province within North America.
\end{abstract}

\end{frontmatter}

% ------------------------------------------------------------------------ %%

 % BEGIN ARTICLE

% ------------------------------------------------------------------------ %%
\linenumbers
\section{Introduction}
It has long been recognized that the sub-continental lithospheric mantle plays an important role in stabilizing tectonic plates against disruption.  The stability of the lithospheric mantle beneath continents is primarily a function of its thermal structure and buoyancy modification by melt extraction \citep{lee2001,lenardic2003longevity}.  It is also widely recognized that stable continental lithosphere may be destabilized by a disruption of its buoyancy and thermal structure, for example, associated with thickening and removal of dense lithospheric mantle in ``delamination'' events \citep{elkinstanton2005continental,molnar2004test} as suggested in the Sierra Nevada \citep{ZandtCarrigan93} and elsewhere in the western US \citep{west2009vertical,schmandt2010,levander11}.  We explore an alternative process by which stable tectonic regions may be destabilized, namely, thermal and chemical modification during infiltration of melts into the lithospheric column. The potential importance of this ``rejuvenation'' process has been described in several peridotite massifs \citep{bodinier2008,  Leroux:2008p710, soustelle2009, Marchesi2010} as the migration of a ``refertilization'' front that is driven by chemical and physical disequilibrium between infiltrating fluids and the surrounding lithosphere. In this work, we use the terms refertilization and rejuvenation, with emphasis on the chemical and mechanical implications, respectively, and will explore the regional-scale geodynamic factors that control the process.  Unless otherwise specified, we use the term lithosphere to signify the thermal boundary layer, although we will occasionally specify the compositionally-distinct chemical boundary layer when discussing xenolith data \citep{Fischer:2010}.

This work is motivated by observations pertinent to the structure and Cenozoic evolution of the Colorado Plateau that suggest rejuvenation by interaction with melts.  First, the Colorado Plateau has been minimally affected by deformation and magmatism during Cenozoic time, but there is evidence for active deformation at its margins (Figure 1; \citet{Berglund2012,Kreemer2010}).  The internal strength of the Colorado Plateau has been attributed to a greater degree of iron-depletion, evidenced by mantle xenoliths with iron-magnesium ratios that are similar to those in Archean cratons and are distinct from those in surrounding regions \citep{alibert1990,smith2000insights,lee2001,roy2009}. The margins of the plateau however show evidence for Cenozoic deformation and have been affected by a slow encroachment of magmatic activity toward the plateau-interior since 40 Ma \citep{roy2009,wenrich1995spatial}.  Seismic observations reveal slower wave speeds in the mantle at the margins of the plateau \citep{schmandt2010,humphreys1994western,goes2002,gao2004upper,sine2008mantle,Li_LAB_2007, tian09} and higher $V_{p}/V_{s}$ ratios at the margins relative to the plateau interior \citep{schmandt2010}.  Average surface heat flow for the margins of the plateau are comparable to values in the Basin and Range province (85-102 mW/m$^{2}$), whereas average surface heat flow in the interior is lower \citep[55-69 mW/m$^{2}$;][]{eggleston_reiter,swanberg_morgan}.  

Higher seismic wavespeeds and lower heat flow in the interior of the Colorado Plateau, together with the composition and thermobarometry of xenoliths entrained in Cenozoic volcanics, indicate thick thermal and chemical lithosphere \citep{smith2000insights,lee2001,roy2009}.  Lower seismic wavespeeds, higher heat flow, and greater magmatism and seismicity at the margins of the plateau relative to the interior suggest that the lithosphere beneath the margins of the plateau is distinct from that of the interior.  Sharp lateral gradients in seismic wavespeeds between the NW plateau margin (Utah Transition Zone) and the interior of the plateau \citep{schmandt2010,gao2004upper,sine2008mantle,Li_LAB_2007} have prompted the suggestion that the margins of the plateau have been mechanically thinned or removed \citep[e.g.,][]{schmandt2010,vanwijk10,levander11}.  We argue that these observations, together with rock uplift and stratigraphic data, are also consistent with thermal and chemical modification by melt-infiltration at the margins of the plateau (Figure 1).  Mechanical erosion or thinning and removal could be part of this modification, but we focus on the thermo-chemical aspects of melt-infiltration as an end-member on a spectrum of processes. 

Specifically, we propose a process that is controlled by interactions at the lithosphere-asthenosphere boundary as North America moves with respect to the underlying asthenosphere. The removal of the subducted Farallon plate from beneath North America in middle-Tertiary time provides an important physical perturbation to the lithosphere-asthenosphere system \citep{humphreys2003}, that essentially triggers the melt-infiltration process explored here.  We present two lines of evidence for melt-infiltration as a driving mechanism for lithospheric rejuvenation and destabilization at the margins of regions of thick lithosphere.  First, a simple numerical formulation of two-phase flow shows that melts and metasomatic fluids would preferentially infiltrate and thus modify regions with a gradient in plate thickness, for example at the margins of regions of thick lithosphere.  We predict that infiltration of melt into the margins of regions of thick lithosphere is asymmetric, with enhanced infiltration at the ``upwind" margins of the protrusion, with limited to no infiltration at the ``downwind" side.  This explains first-order patterns in the rates and asymmetry of magmatic encroachment at the Colorado Plateau. Second, based on experimental observations described below, we infer that the regions of greatest melt infiltration are likely to be regions of enhanced viscosity reduction and strain concentration. In the models presented here, we do not explicitly include coupling of deformation and fluid segregation and organization, but instead present heuristic arguments for the importance of this stress-driven feedback based on numerical and laboratory experiments.  Finally, we speculate about the general importance of the melt-infiltration process explored here and its role in the regional evolution of the western US. 

%=================================================
\section{Stress-driven melt-segregation and melt-infiltration}

The dynamic organization of melt during deformation is an area of active field, laboratory and theoretical research.  In this work we make a distinction between stress-driven melt-segregation (SDS), which is the re-distribution of melt at the grain-scale during deformation of a molten aggregate, and the large-scale motion of melt relative to the surrounding rock, which we refer to as melt infiltration and extraction.  The characteristic length scale for distinguishing these processes of melt-rock interaction is the compaction length, $\delta_c$, which depends on the solid and fluid viscosities and the permeability, $\delta_c = \sqrt{\kappa(\zeta+4/3\eta)/\mu}$.  For typical viscosity contrasts between melt and the solid (e.g., melt shear viscosity $\mu=1 $ Pa s, and surrounding rock bulk and shear viscosity of $\zeta$ and $\eta=10^{21}$ to $10^{24}$ Pa s; see parameters in Table 1), and reasonable rock permeabilities ($\kappa=10^{-15}$ to $10^{-12}$ m$^2$), the compaction length scale for the upper mantle is on the order of $\delta_c=10^1$ to $10^4$ m \citep{mckenzie1984}.  In the following we divide our discussion into melt-infiltration and extraction over regional tectonic scales (much larger than the compaction length) and stress-driven melt segregation processes that are relevant at the grain-scale up to the compaction length.  Although this division in terms of the compaction length scale is a convenient way to categorize the physics of magma migration, we note that interaction between processes across these two length scales will play an important role in melt migration through the asthenosphere and deeper portions of the lithosphere.

%=================================================
\subsection{Idealized model of spatially-variable melt-infiltration and extraction}

On length scales that are large compared to the compaction length, we expect pressure-gradients within the deforming lithosphere-asthenosphere system to drive melt-infiltration into the lithosphere and thus control the distribution of magmatic activity at the surface.  To first order, the interactions between melt and a deformable solid matrix may be treated as a coupled problem involving flow of two interprenetrating fluids \citep{mckenzie1984,spieg93,spieg2003} consisting of a low-viscosity phase such as melt and a high viscosity phase representing the matrix.  Analytic solutions for the two-phase flow problem exist in idealized cases \citep{mckenzie1984,spieg93} and if porosity gradients are nearly zero, relative motion between the solid and melt, or the melt extraction rate, is primarily driven by pressure gradients 
\begin{linenomath*}
\begin{equation}
-\frac{ \mu\phi}{\kappa_{\phi}} (\mathbf{v}-\mathbf{u})=\mathbf{\nabla} P - \rho_{f} \mathbf{g}
\end{equation} 
\end{linenomath*}
where we explicitly note that permeability may be a function of porosity, $\mathbf{u}$ is the melt velocity, and $\mathbf{v}$ is solid velocity \citep{mckenzie1984,spieg93}.  Although these assumptions are simplistic, they define a first-order approach to understanding how buoyancy-driven flow of melt will interact with dynamic pressure gradients within the asthenosphere and lithosphere to drive spatially-variable melt extraction rates.

%=================================================
\subsubsection{Melt organization due to variable lithosphere thickness}

We first consider a scenario in which a region of variable-thickness lithosphere (represented by a high-viscosity Newtonian fluid, $\eta = \eta_{l}$) protrudes into and moves relative to an asthenospheric ``wind" (of a lower viscosity Newtonian fluid, $\eta = \eta_{a}$).  The Couette-style shear between the plate and the underlying mantle is modified by flow around the lithospheric protrusion(s) (Figure 2a and b).  This flow field may represent, for example, either eastward-driven asthenosphere or westward migration of the North American plate relative to the asthenospheric mantle.  The keel-like protrusion represents a region of thicker than average lithosphere, such as the Colorado Plateau. If the viscosity contrast between the lithosphere and asthenosphere is large (in these calculations $\eta_{l}/\eta_{a} = 10^{3-4}$;  Figure 2c), the keel-like protrusion will not deform greatly and the asthenospheric flow produces a ``pressure shadow" effect within the keel: lower dynamic pressure on the upwind side of the keel and higher on the downwind side (Figure 3a). A similar effect is produced for 2D models of semicircular and periodic sinusoidal ridges at the base of the plate (Figure 3b).   

For simplicity, we ignore processes that generate partial-melt in the lithosphere-asthenosphere system, but assume instead that a small, uniform melt-fraction is present everywhere within the model domain.  In this case, equation (1) may be used to determine the melt extraction field: relative motion between the melt and the solid is driven by the competition between dynamic pressure gradients and buoyancy (right-hand side of equation 1).  In the absence of dynamic pressure gradients (no deformation in the lithosphere and asthenosphere), the buoyancy term will drive upward motion of melt relative to the solid.  If horizontal dynamic pressure gradients are sufficiently large, then melt motion relative to the solid will be more complicated. The geometry of the melt-extraction rate field is illustrated by calculating forward-streamlines for tracers that originate at a fixed depth, taken to be below the lithosphere protrusion(s) in our calculations (Figure 3). The effects of dynamic pressure gradients may be seen as the departure of the streamlines from vertical (buoyancy-dominated) orientation (Figure 3).  

Asymmetry in the dynamic pressure gradients drive an asymmetry in melt-extraction on the upwind and downwind sides of the 3D hemispherical keel (Figure 3a) and also in the 2D models of semicircular and periodic ridges (Figure 3b).  On the upwind side, asthenospheric flow diverges around the protrusions and the upwind sector of the keel or ridge has lower dynamic pressure than the downwind sector (Figures 2 \& 3).  The dynamic pressure field thus focusses melt streamlines into the upwind sector of the protrusions and pushes them away from the downwind sector of the protrusions, producing a pronounced upwind-downwind asymmetry due to asthenospheric flow (Figure 3).

%=================================================
\subsubsection{Thermal effects of asthenosphere flow and melt-infiltration}

The models above solve the coupled Navier-Stokes and heat equations for a lithospheric region of high viscosity (hemisphere or ridges) that initially has a cool linear geotherm and an asthenospheric fluid with fixed, higher uniform temperature ($1200\,^{\circ}{\rm C}$).  The flow is accompanied by warming of the lithosphere, due to conduction \citep{roy2009} and advective heating by asthenospheric flow. We assume that the viscosities of the lithosphere and asthenosphere are temperature-dependent (assuming $\eta_{l}=\eta_{0}\exp(-E(T-T_{0}))$, e.g., \citet{zhong2000}; parameters in Table 1) so that as the lithosphere warms, it weakens and deforms more easily.  The deformation and dynamic pressure fields will evolve due to heating (Figure 4) and drive changes in the pattern of melt-extraction (Figure 4).  At early times during the simulations the pattern of melt-extraction is asymmetric, with melt drawn into the upwind side and pushed out of the downwind side of the protrusions.  Interestingly, the surface pattern of asymmetric melt-extraction persists at the surface over timescales of $10^{7}$ years for lithospheric protrusions with linear dimensions of $10^{2}$ km across. At depth, however, warming and viscosity-reduction in the lithosphere causes diminishing gradients in dynamic pressure, so that melt-streamlines become increasingly vertical through time (Figure 4).  

On the upwind side of the protrusion advective heating by the melt-infiltration enhances weakening of the lithosphere.  At steady-state, assuming a constant melt-extraction field, advective heat transport by melt raises temperatures above background by around $50-100\,^{\circ}{\rm C}$ \citep[see also ][]{schmeling12}.  The additional heating would provide an important feedback between melt-infiltration and deformation, with melt-infiltration reducing plate viscosity by and additional $\sim40 \%$.  The asymmetry in melt-extraction will drive an asymmetry in destabilization, namely the upwind side of the protruding lithosphere experiences this enhanced viscosity reduction relative to the melt-free downwind sector.  We predict, therefore, that lithospheric rejuvenation driven by melt-infiltration will be enhanced on the upwind sides of lithospheric protrusions.  It is important to note that a more complete treatment of the problem will require not only this thermal feedback between melt-infiltration and plate viscosity, but also important feedbacks between melt-segregation and deformation observed in laboratory experiments at (and below) compaction length scales.  

%=================================================
\subsection{Laboratory observations of melt and rheology at the meso-scale}
\label{sec:experiments}
In deformation experiments on partially molten rocks, samples with initially homogeneous melt distributions undergo a deformation-driven transition where melt organizes into networks of connected, anastomosing shear zones \citep[e.g.,][]{Kohlstedt:2009}. 
This process, referred to as stress-driven segregation (SDS), may provide an important feedback between melt distribution, transport, and rheology that controls the geometry and distribution of melt. Laboratory experiments suggest that SDS is more likely in higher-strain regions of the mantle, but the degree of strain and the feedbacks in an open system (and thus the volume in which the process occurs in the mantle) are not yet known. 
The notion of the compaction length, the characteristic length scale of coupling between solid matrix deformation and fluid flow, has been useful in predicting behavior in experiments \citep[e.g.,][]{holtzman2003stressdriven}, and is consistent with the length scale of observed organized-melt structures when extrapolated from laboratory conditions to mantle and lower crust \citep[e.g.,][]{holtzman2007stressdriven}.  For a region with homogeneous melt distribution the compaction length is uniform, but as melt organizes and segregates, at length scales smaller than that initial compaction length, $\delta_c$ becomes spatially heterogeneous. Because the melt organization will produce anisotropic permeability and viscosity structures, the effective compaction length would also be anisotropic.  While the characteristic spacing of the largest melt-rich features is predicted to be smaller than the compaction length, the linear dimensions of the volume in which the process occurs could be much larger than the initial compaction length.  

In experiments starting from a homogeneous melt distribution, a degree of strain is required to segregate the melt to a point at which it begins to affect the rheological and melt-transport properties. 
This critical degree of strain, $\gamma_c$, shows some sensitivity to stress, and is thus likely to be a function of thermodynamic conditions. While previous work focused on the large bands that emerge by about a shear strain of $\gamma = \pm 1$ as an indicator of $\gamma_c$ \citep{holtzman2007stressdriven,Katz:2006}, \citet{holtzman12} demonstrate that the smallest scale bands form at much lower strains but cause a clear and immediate reduction of the effective viscosity.  In \citet{holtzman2007stressdriven} it was proposed that the weakening is associated with high mobility of melt through a network of connected bands. The melt must form connected pathways, even if transiently, in order to reorganize easily to concentrate strain on low angle bands. Thus, the effective permeability is inferred to increase at the onset of weakening. 

When extrapolating to an open system with thermal and compositional gradients, rapid melt transport can produce chemical and thermal disequilibrium between melt and host rock \citep{Kohlstedt:2009}. The effects of chemical disequilibrium under gravity-driven melt flow has been well studied, especially in near-adiabatic conditions \citep[e.g.,][]{daines94, kelemen1995}. A method for exploring the coupling of SDS and chemical disequilibrium has demonstrated a strong coupling between the two, resulting in greatly enhanced infiltration \citep{King:2011p6141}.  Calculations in \citet{holtzman12}, based on a model for permeability and its variation with degree of strain, suggest that SDS will enhance both the thermal and chemical disequilibrium in the rock volume. The potential feedbacks between thermal and chemical interactions and the microstructure of the rock (and thus the permeability) are not included in our model. While these disequilibrium processes are complex and difficult to predict, especially for open systems, we surmise that permeability would generally be enhanced in regions of SDS in the mantle. 

To determine if and where the SDS process may occur in the mantle requires two approaches. The first is detailed scaling of the physics from laboratory- to mantle conditions \citep[e.g.,][]{Takei:2009p2204,King:2011p6140,butler2012}. The second approach is to develop predictive tests of dynamical consequences and observational signatures, such as the seismic velocity structure \citep[e.g.,][]{Holtzman:2010,Takei:2009p1421} or electrical conductivity structure \citep{wannamaker08}.  In our discussion below, we follow the second approach by combining insights from numerical models of melt extraction and experimental observations to hypothesize on the role of melt-infiltration and its consequences for the Colorado Plateau and the western US. 

%================================================= 
\section{Discussion}

The models above, together with experimental observations of the behavior of molten aggregates, suggest that melt-infiltration and stress-driven melt segregation must play an important role in the interaction of partial melts with the base of the lithosphere.  To first-order, the geometry of the lithosphere-asthenosphere boundary controls dynamic pressure gradients due to deformation, which in turn control the spatial distribution of melt-extraction and observed magmatism.  Regions with lateral variations in lithosphere thickness will be characterized by complexity in the relative motion between the asthenosphere and lithosphere, as explored here for keel- and ridge-like lithospheric protrusions.  Melt-infiltration preferentially occurs on the upwind side of regions of thick lithosphere, resulting in enhanced weakening here and asymmetric thermal and chemical modification/rejuvenation of the lithosphere.  Simultaneously high strain rates and melt-extraction rates in deforming parts of the protrusion will likely undergo SDS and form melt-rich, anastomosing shear zones, as observed in experiments.  Although neither the model calculations nor the experimental observations discussed above address the generation of partial-melt in the upper mantle, the Cenozoic evolution of the western US is characterized by a number of processes that can generate partial melting in the upper mantle beneath North America.  We speculate, therefore, that SDS and melt-infiltration at the base of the lithosphere may profoundly control the distribution of magmatism and intra-plate deformation in the western US.

%================================================= 
\subsection{Application to the Colorado Plateau}

On a regional scale the spatial pattern of Cenozoic magmatism in the western US exhibits a magmatic gap across the Colorado Plateau, with the encroachment of the onset of magmatism inward toward the interior of the plateau on its NW, SW, and SE margins, with a distinct lack of encroachment on the NE margin \citep{roy2009}.  We argue that this pattern of magmatism may have occurred through a range of processes, both thermal and chemical, associated with melt-rock interaction at the lithosphere-asthenosphere boundary.  Our models suggest that, among other factors, a protruding region of thicker lithosphere and accompanying regional scale dynamic pressure gradients may be responsible for organizing the surface distribution of magmatism surrounding and within the Colorado Plateau.  In the context of specific observations at the Colorado Plateau, the melt-infiltration mechanism discussed here provides a unifying framework within which a number of disparate observations may be explained. 

\subsubsection{Spatio-temporal patterns of magmatism.}  We assume that the middle-Tertiary removal of the Farallon plate subjected the North America to a basal thermal perturbation \citep{roy2009,humphreys2003} and that relative motion between North America and the underlying asthenosphere has been steady since this time \citep{silver2002mantle}. In our models, the relative motion of North America over the underlying asthenosphere is the model is $\sim5$ cm/yr, comparable to rates inferred from the correlation between GPS-derived surface motions and seismic anisotropy  \citep{silver2002mantle}.  Additionally, upwelling flow in the asthenosphere in response to the sinking Farallon plate would drive both long-wavelength dynamic rock uplift \citep{moucha2008mantle,liugurnis10} and generate accompanying partial melt in the asthenosphere beneath the western US.  The voluminous and regionally-extensive middle-Tertiary ignimbrite flareup, for example, has been attributed to the removal of the Farallon slab and subsequent interaction between asthenospheric and lithospheric partial melts \citep{humphreys2003}. Thermal expansion due to conductive relaxation of isotherms following the removal of the Farallon slab is sufficient to explain a large fraction of the rock uplift of the Colorado Plateau and the rate of migration of the isotherms is comparable to (though slightly slower than) the observed slow rates of magmatic encroachment (3-6 km/my) onto the plateau \citep{roy2009}.  

The models presented here refine the findings in \citet{roy2009}, where the Colorado Plateau is represented by a region where a deep ÔkeelÕ of lithosphere protrudes into the asthenosphere, around which the asthenosphere must flow. Our models offer an explanation for the observed asymmetry in magmatic encroachment surrounding the Colorado Plateau, which was not previously explained in the simpler conductive models of \citet{roy2009}.  Specifically, we note that in a frame of reference stationary with respect to North America, the asthenosphere below the plate is inferred to move in a roughly SW-NE direction (Figure 1).  Consistent with our models, the highest rates of magmatic encroachment onto the plateau ($\sim6$ km/my) are observed in the upwind sector of the plateau (its SW quadrant), whereas magmatic encroachment occurs at a slower rate on the NW and SE quadrants \citep{roy2009}.  In contrast, the downwind sector (its NE quadrant, adjacent to the San Juan volcanic field and northern Rio Grande rift) shows a distinct lack of magmatic encroachment \citep{roy2009}. 

In our models, the expected rate at which the zone of melt focusing migrates into the upwind side of the protruding lithosphere is a function of time, ranging from values as high as $6$ km/my to slower rates $\sim2-3$ km/my (Figure 5).  These rates of migration of the magmatic zone are comparable to observed magmatic encroachment rates estimated by \citet{roy2009} at the SW (6.3 km/my), NW (4.0 km/my) and SE (3.3 km/my) from ages of volcanic rocks.  The predicted migration rates in our models are slightly higher than those predicted by conductive models alone \citep{roy2009}, as they are enhanced by advective heating due to asthenosphere flow.  

In contrast, at the downwind sector (the NE quadrant) the distinct lack of magmatic encroachment \citep{roy2009} is consistent with higher dynamic pressures and therefore melt exclusion from this part of the plateau (Figure 5). Alternatively, the lack of magmatic encroachment on the NE margin of the Colorado Plateau may be due to the lower-magnitude of extension in the northern Rio Grande rift \citep{chapin94} which results in smaller contrast in lithosphere thickness across the plateau margin.  Our models predict a long-lived pattern of melt-exclusion from the downwind sector of the plateau and protracted magmatism immediately outside the margin of the plateau (Figures 4 \& 5).  These predictions are consistent with the voluminous and long-lived magmatic centers of the middle-Tertiary to Neogene San Juan Volcanic Field \citep{lipman1991} located NE of the plateau, outside its physiographic boundary. The northern margin of the plateau (adjacent to the Archean Wyoming Province) lacks magmatism and the relatively high upper mantle seismic wave speeds of the plateau interior continue into the Wyoming Province to the north.  The transition into the Wyoming Province is therefore consistent with a more uniform lithosphere thickness.

\subsubsection{Properties of the Colorado Plateau lithosphere.} 
Central to the idea of middle-Tertiary to Present melt-infiltration and magma migration into the margins of the Colorado Plateau is the excess thickness of the lithosphere beneath the Colorado Plateau relative to its surroundings.  In the models explored here, asthenospheric flow around protruding regions of thicker lithosphere sets up the key dynamic pressure gradients that are responsible for organizing melt-migration.  Evidence for thicker lithosphere beneath the Colorado Plateau relative to its surroundings is provided by a range of seismic observations \citep{schmandt2010,lin_ritzwoller2011,sine2008mantle,west2004crust,levandermiller12} and by xenolith data \citep{smith2000insights,lee2001}.  

The magmatic infiltration process described here is further supported by the distinct geologic and geophysical contrast between the margins and the interior of the Colorado Plateau.  The asthenosphere beneath the Rio Grande rift and the Basin and Range exhibit extremely low shear velocities \citep[e.g.,][]{west2004crust,Yang2008, RauForsyth11}, suggestive of ubiquitous partial melt surrounding the plateau.  Detailed seismic imaging clearly demonstrates that these low relative velocities extend significantly under the margins of the plateau, well inside the outer plateau's physiographic boundary \citep{schmandt2010, sine2008mantle}.  In contrast, the interior of the plateau is seismically faster, with lower $V_p/V_s$ ratios, than its margins \citep{schmandt2010,humphreys1994western,goes2002,gao2004upper,sine2008mantle,Li_LAB_2007}.  The plateau margins are also characterized by higher average surface heatflow  \citep{eggleston_reiter,swanberg_morgan} and enhanced electrical conductivity in the upper mantle \citep{wannamaker08}. The spatial correlation between the regions of magmatism at the plateau margins and the observed geophysical structure (Figure 1), strongly suggests that the margins of the plateau have been modified by melt-infiltration and migration via processes akin to those explored in our models.

Within the higher-velocity plateau interior, recent studies image a localized region of fast seismic velocities to a depth of over 200 km adjacent to the Utah Transition Zone \citep{schmandt2010,sine2008mantle,levander11}, interpreted as a region of lithospheric delamination or a ``drip'' at the base of the plate \citep{levander11}.  There are uncertainties with this interpretation, in particular with respect to the apparent lithosphere-asthenosphere boundary imaged near 150 km depth continuously across the drip region \citep{levandermiller12}.  However, given that delamination is a plausible explanation for the deep feature, we note that the melt-infiltration process may promote lithospheric destabilization and downwelling.  In particular, the feedback mechanisms between melt segregation, viscosity reduction, and deformation described in section~\ref{sec:experiments} provides a natural and self-consistent source for the localized weakening that is required to drive delamination in the absence of lithospheric thickening. In this scenario, melt infiltration, migration, and associated magmatism drive the delamination process, rather than vice-versa. 

The contrast between the plateau interior and its margins is also evident in the petrology of mantle xenoliths from the Colorado Plateau region.  Xenolith populations show a predominance of pyroxenite lithologies at the plateau margins relative to the interior \citep{roy2005}, with evidence for melt-related metasomatic alteration of lherzolite into pyroxenite via interaction with fluids of silicate and carbonatite composition \citep{porreca2006pyroxenite}.  These observations raise the possibility that the seismically- and petrologically-distinct margins of the Colorado Plateau are modified by the infiltration of melts, through a combination of thermal effects (as explored in our models) and also chemical changes, leading to a thermo-chemical corrosion of the lithosphere.

\subsubsection{Relationship between magmatism and strain rate.} The melt-related lithospheric rejuvenation process proposed here involves \emph{in situ} modification of the margins of the region of thick lithosphere and, to the extent that it is controlled by the shape of the base of North America, it must also depend on the history of extension surrounding the plateau (Figure 6a).  Although in our models the lithosphere is not undergoing concurrent extension and warming, we note intriguing relationships between the pattern of magmatic encroachment and present-day strain-rates.  

In particular, if melt-segregation and extraction is a strong function of the SDS feedback between matrix viscosity, permeability, and strain-rate (see \ref{sec:experiments}), then we would expect that melt-segregation into planar bands and melt-migration and extraction along melt bands will be most efficient in regions of highest strain rate (Figure 6b, c, \&d).  The present-day strain rate field across the Colorado Plateau is characterized by limited extension at the plateau margins, with the exception of the NW margin at the Utah Transition Zone between the plateau and the Great Basin \citep{Kreemer2010,Berglund2012}. Current strain-rates are highest at this margin and are spatially correlated with seismicity and magmatism.  We would expect that SDS would most efficiently segregate melt along the NW margin, allowing channelized melt-transport.  This focusing effectively \emph{reduces} the percolative corrosion at greater depths predicted in the models, by enhancing the transport upwards along the LAB (Figure 6c), focusing and enhancing thermochemical corrosion at shallower depths because of the increased degree of thermal and chemical disequilibrium between the melt and the lithospheric mantle.  This effect may explain the dramatic gradients in seismic velocity observed at the NW margin of the plateau \citep{schmandt2010}, and the reduced magmatic encroachment in this quadrant compared to the SW (upwind) margin (4.0 km/my as opposed to 6.3 km/my \citep{roy2009}).  Magneto-telluric measurements of the polarization anisotropy of electrical resistivity within the Utah Transition Zone are consistent with the presence of N-S oriented partial melt lenses, parallel to the plateau margin \citep{wannamaker08}, further supporting the role of SDS in this region. 

The broad range of observations discussed above, including the patterns of magmatism, heatflow, seismic and electrical structure, and xenolith data, suggest that the margins and likely the base of the thicker Colorado Plateau lithosphere has undergone variable thermo-chemical modification, driven by melt-infiltration.  The distribution and rates of melt-infiltration are likely modulated by gradients in lithospheric thickness and strain-rates at the lithosphere-asthenosphere boundary.   In the following we consider the implications of this process within the broader context of the western US and suggest that the thermo-chemical corrosion mechanism proposed here may be regionally significant. 

%================================================= 
\subsection{Application to the Cenozoic evolution of the western US and beyond.}

Following removal of the Farallon slab and the subsequent introduction of fresh asthenosphere material beneath the western US  \citep{humphreys2003}, we speculate that melt-related rejuvenation may have occurred on a regional scale at the margins and at the base of regions of thicker lithosphere.  Introduction of hot asthenosphere (e.g., ``sub-EPR asthenosphere" in Figure 6a) triggers a spatially-variable melt-infiltration and rejuvenation of the North American plate.  Specifically, the melt-infiltration and thermo-chemical modification inferred for the margins of the Colorado Plateau may have also affected the previously thick lithosphere of the now-extended Basin and Range Province and may have variably affected the thicker Wyoming craton to the north.  Evidence that interaction with a LILE-rich melt or fluid resulted in re-enrichment of incompatible trace elements in the lower lithosphere and within the Great Falls Tectonic Zone at the margin of the Wyoming craton \citep{carlson2004} suggests that metasomatism may play a key role in the evolution of this region. 

In this view, the continuous region of present-day higher seismic wavespeeds encompassing both the interior of the Colorado Plateau and Wyoming Province \citep[e.g.,][]{tian09,schmandt2010,levander11,west2004crust,lin_ritzwoller2011,xueallen2010,wagner10, sigloch11}, represents thicker portions of the lithosphere that has been only partially affected by melt-infiltration (Figure 6c).  In contrast, the thinner, seismically slow lithosphere of the Basin and Range province may have been more completely rejuvenated and destabilized by melt-infiltration.  The importance of partial melt within the seismically slow Basin and Range province is underscored by low shear wave speeds \citep{RauForsyth11} and by higher $V_{p}/V_{s}$ ratios \citep{schmandt2010}.

Thinking beyond the western US, we note that recent work in several localities, including the Ordos block of the North China craton \citep{menzies2007,kusky2007tectonic}, the eastern part of the Sino-Korean craton \citep{reilly2001are}, and the Tanzania craton \citep{wolbern12} has suggested that within tectonically stable plate interiors, melt-infiltration and its effects can lead to rejuvenation and destabilization.  These studies infer that melt-infiltration and its thermo-chemical consequences can transform previously stable, cratonic provinces into tectonized, orogenic zones.  In the North China craton, for example, interaction of the lithosphere with infiltrating melts is inferred to play a key role in the evolution of the eastern part of the craton from a region with a cool geotherm and a thick, highly magnesian keel into a tectonized zone characterized by present-day low upper-mantle seismic velocities and thin lithosphere \citep{menzies2007}. In Tanzania, a mid-lithosphere discontinuity imaged by S-wave receiver functions forms the upper boundary of a zone of reduced shear wave speeds in the lower lithosphere \citep{wolbern12}.  In this study, the lower lithosphere is inferred to have been modified by melt-infiltration and accompanying alteration of mineral assemblages \citep{wolbern12}, in a process akin to that inferred for the Wyoming Province \citep{carlson2004}.  We speculate, therefore, that melt-infiltration and thermo-chemical alteration may play an important role in the general evolution of continental interiors, particularly at the margins of thicker cratonic lithosphere.
%=================================================
\section{Conclusions}
%================================================= 
The well-studied Colorado Plateau region provides a unique laboratory for investigating processes by which stable continental interiors may be destabilized, namely, thermal and chemical modification during infiltration of melts or metasomatic fluids into the lithosphere.  To first-order, the geometry of the lithosphere-asthenosphere boundary controls dynamic pressure gradients due to deformation, which in turn control the spatial distribution of melt-extraction and the observed distribution of magmatism.  For a region of thicker lithosphere that protrudes into the ``mantle wind", such as the Colorado Plateau, the dynamic pressures are lowest on the upwind side and highest on the downwind side. In this scenario, the highest rates of melt-infiltration and magmatic encroachment occur in the upwind sector of the plateau (its SW quadrant), whereas magmatic encroachment occurs at a slower rate on the NW and SE quadrants.  In contrast, the downwind sector (NE margin, adjacent to the San Juan volcanic field and northern Rio Grande rift) shows a distinct lack of magmatic encroachment, consistent with our melt-extraction models.  The melt-infiltration and rejuvenation process explored here provides a unifying framework for explaining both the observed asymmetry in magmatic patterns and also the observed geophysical contrasts between the plateau interior and its margins. Stress-driven melt segregation may play an important role in enhanced melt transport at the highest strain-rate NW margin of the plateau.  In the Utah Transition Zone for example, sharp lateral seismic velocity and electrical conductivity gradients in the upper mantle may be due to the presence of segregated melt in oriented planar bands.  

We speculate that destabilization and rejuvenation of the lithosphere by melt-infiltration may have played a regionally important role in the Cenozoic evolution of the western US.  For example, geochemical observations in the Wyoming Province suggest that the lower lithosphere in the region may have been partially modified by interaction with melts and metasomatic fluids.  The seismically-fast interior of the Colorado Plateau and the adjacent Wyoming craton to the north may be portions of thicker North American lithosphere that have been less affected by melt-infiltration.  In contrast, the margins of the Colorado Plateau and the thinned lithosphere of the Great Basin represent rejuvenated portions of the plate, both thermally and chemically modified by interaction with melts.  Feedback between melt segregation, viscosity reduction, and deformation provides a natural and self-consistent means of weakening the lithosphere in the western US and may have hastened its thinning by a combination of extension and convective removal.  Melt-infiltration and accompanying rejuvenation may have played an important role in the evolution of other continental interiors, such as the Sino-Korean, North China and Tanzania cratons, suggesting its general importance for the of continents.

%
%  ACKNOWLEDGMENTS

\section{Acknowledgments}
We thank A. Ringler for early FEM models and B. Schmandt and E. Humphreys for providing a slice through their tomography model for Figure 1.  We also thank M. Spiegelman and T. Plank for fruitful conversations on magma and the Colorado Plateau.

%% ------------------------------------------------------------------------ %%
%%  REFERENCE LIST AND TEXT CITATIONS

%\bibliography{MyRefs}{}

\begin{thebibliography}{69}
\expandafter\ifx\csname natexlab\endcsname\relax\def\natexlab#1{#1}\fi
\expandafter\ifx\csname url\endcsname\relax
  \def\url#1{\texttt{#1}}\fi
\expandafter\ifx\csname urlprefix\endcsname\relax\def\urlprefix{URL }\fi
\providecommand{\eprint}[2][]{\url{#2}}
\providecommand{\bibinfo}[2]{#2}
\ifx\xfnm\relax \def\xfnm[#1]{\unskip,\space#1}\fi
%Type = Article
\bibitem[{Alibert(1990)}]{alibert1990}
\bibinfo{author}{Alibert, C.}, \bibinfo{year}{1990}.
\newblock \bibinfo{title}{Peridotite xenoliths from the western grand canyon
  and the thumb: a probe into the subcontinental mantle of the colorado
  plateau}.
\newblock \bibinfo{journal}{Journal of Geophysical Research}
  \bibinfo{volume}{99}, \bibinfo{pages}{21605--21620}.
%Type = Article
\bibitem[{Berglund et~al.(2012)Berglund, Sheehan, Murray, Roy, Lowry, Nerem and
  Blume}]{Berglund2012}
\bibinfo{author}{Berglund, H.T.}, \bibinfo{author}{Sheehan, A.F.},
  \bibinfo{author}{Murray, M.H.}, \bibinfo{author}{Roy, M.},
  \bibinfo{author}{Lowry, A.R.}, \bibinfo{author}{Nerem, R.S.},
  \bibinfo{author}{Blume, F.}, \bibinfo{year}{2012}.
\newblock \bibinfo{title}{Distributed deformation across the rio grande rift,
  great plains, and colorado plateau}.
\newblock \bibinfo{journal}{Geology} \bibinfo{volume}{40},
  \bibinfo{pages}{23--26}.
%Type = Article
\bibitem[{Bodinier et~al.(2008)Bodinier, Garrido, Chanefo, Brugier and
  Gervilla}]{bodinier2008}
\bibinfo{author}{Bodinier, J.L.}, \bibinfo{author}{Garrido, C.J.},
  \bibinfo{author}{Chanefo, I.}, \bibinfo{author}{Brugier, O.},
  \bibinfo{author}{Gervilla, F.}, \bibinfo{year}{2008}.
\newblock \bibinfo{title}{Origin of pyroxenite-peridotite veined mantle by
  refertilization reactions: Evidence from the ronda peridotite (southern
  spain)}.
\newblock \bibinfo{journal}{Journal of Petrology} \bibinfo{volume}{49}.
%Type = Article
\bibitem[{Butler(2012)}]{butler2012}
\bibinfo{author}{Butler, S.L.}, \bibinfo{year}{2012}.
\newblock \bibinfo{title}{Numerical models of shear-induced melt band formation
  with anisotropic matrix viscosity}.
\newblock \bibinfo{journal}{Physics of The Earth and Planetary Interiors}
  \bibinfo{volume}{200-201}, \bibinfo{pages}{28--36}.
%Type = Article
\bibitem[{Carlson et~al.(2004)Carlson, Irving, Schulzec and Jr}]{carlson2004}
\bibinfo{author}{Carlson, R.W.}, \bibinfo{author}{Irving, A.J.},
  \bibinfo{author}{Schulzec, D.J.}, \bibinfo{author}{Jr, B.C.H.},
  \bibinfo{year}{2004}.
\newblock \bibinfo{title}{Timing of precambrian melt depletion and phanerozoic
  refertilization events in the lithospheric mantle of the wyoming craton and
  adjacent central plains orogen}.
\newblock \bibinfo{journal}{Lithos} \bibinfo{volume}{77}.
%Type = Book
\bibitem[{Chapin and Cather(1994)}]{chapin94}
\bibinfo{author}{Chapin, C.E.}, \bibinfo{author}{Cather, S.M.},
  \bibinfo{year}{1994}.
\newblock \bibinfo{title}{Tectonic setting of the axial basins of the northern
  and central Rio Grande rift}. volume \bibinfo{volume}{291} of
  \textit{\bibinfo{series}{Special Paper of the Geological Society of
  America}}.
\newblock \bibinfo{publisher}{Geological Society of America}.
%Type = Article
\bibitem[{Daines and Kohlstedt(1994)}]{daines94}
\bibinfo{author}{Daines, M.}, \bibinfo{author}{Kohlstedt, D.},
  \bibinfo{year}{1994}.
\newblock \bibinfo{title}{The transition from porous to channelized flow due to
  melt/rock reaction during melt migration}.
\newblock \bibinfo{journal}{Geophysical Research Letters} \bibinfo{volume}{21},
  \bibinfo{pages}{145--148}.
%Type = Article
\bibitem[{Eggleston and Reiter(1984)}]{eggleston_reiter}
\bibinfo{author}{Eggleston, R.}, \bibinfo{author}{Reiter, M.},
  \bibinfo{year}{1984}.
\newblock \bibinfo{title}{Terrestrial heat flow estimates from petroleum bottom
  hole temperature data in the colorado plateau and the eastern basin and range
  province}.
\newblock \bibinfo{journal}{Geological Society of America Bulletin}
  \bibinfo{volume}{95}, \bibinfo{pages}{1027--1034}.
%Type = Article
\bibitem[{Elkins-Tanton(2005)}]{elkinstanton2005continental}
\bibinfo{author}{Elkins-Tanton, L.T.}, \bibinfo{year}{2005}.
\newblock \bibinfo{title}{Continental magmatism caused by lithospheric
  delamination}.
\newblock \bibinfo{journal}{Special Paper - Geological Society of America}
  \bibinfo{volume}{388}, \bibinfo{pages}{449--461}.
%Type = Article
\bibitem[{Fischer et~al.(2010)Fischer, Ford, Abt and Rychert}]{Fischer:2010}
\bibinfo{author}{Fischer, K.M.}, \bibinfo{author}{Ford, H.A.},
  \bibinfo{author}{Abt, D.L.}, \bibinfo{author}{Rychert, C.A.},
  \bibinfo{year}{2010}.
\newblock \bibinfo{title}{The lithosphere-asthenosphere boundary}.
\newblock \bibinfo{journal}{Annual Review of Earth and Planetary Science}
  \bibinfo{volume}{38}, \bibinfo{pages}{551--575}.
%Type = Article
\bibitem[{Gao et~al.(2004)Gao, Grand, Baldridge, Wilson, West, Ni and
  Aster}]{gao2004upper}
\bibinfo{author}{Gao, W.}, \bibinfo{author}{Grand, S.},
  \bibinfo{author}{Baldridge, W.S.}, \bibinfo{author}{Wilson, D.},
  \bibinfo{author}{West, M.}, \bibinfo{author}{Ni, J.}, \bibinfo{author}{Aster,
  R.}, \bibinfo{year}{2004}.
\newblock \bibinfo{title}{Upper mantle convection beneath the central {Rio}
  {Grande} rift imaged by {P} and {S} wave tomography}.
\newblock \bibinfo{journal}{Journal of Geophysical Research}
  \bibinfo{volume}{109}.
%Type = Article
\bibitem[{Goes and van~der Lee(2002)}]{goes2002}
\bibinfo{author}{Goes, S.}, \bibinfo{author}{van~der Lee, S.},
  \bibinfo{year}{2002}.
\newblock \bibinfo{title}{Thermal structure of the north american uppermost
  mantle inferred from seismic tomography}.
\newblock \bibinfo{journal}{Journal of Geophysical Research}
  \bibinfo{volume}{107}.
%Type = Article
\bibitem[{Holtzman et~al.(2003)Holtzman, Groebner, Zimmerman, Ginsberg and
  Kohlstedt}]{holtzman2003stressdriven}
\bibinfo{author}{Holtzman, B.K.}, \bibinfo{author}{Groebner, N.J.},
  \bibinfo{author}{Zimmerman, M.E.}, \bibinfo{author}{Ginsberg, S.B.},
  \bibinfo{author}{Kohlstedt, D.L.}, \bibinfo{year}{2003}.
\newblock \bibinfo{title}{Stress-driven melt segregation in partially molten
  rocks}.
\newblock \bibinfo{journal}{Geochem. Geophys. Geosyst} \bibinfo{volume}{4},
  \bibinfo{pages}{26}.
%Type = Article
\bibitem[{Holtzman and Kendall(2010)}]{Holtzman:2010}
\bibinfo{author}{Holtzman, B.K.}, \bibinfo{author}{Kendall, J.M.},
  \bibinfo{year}{2010}.
\newblock \bibinfo{title}{Organized melt, seismic anisotropy, and plate
  boundary lubrication}.
\newblock \bibinfo{journal}{Geochem. Geophys. Geosyst.} \bibinfo{volume}{11},
  \bibinfo{pages}{Q0AB06}.
%Type = Article
\bibitem[{Holtzman et~al.(2012)Holtzman, King and Kohlstedt}]{holtzman12}
\bibinfo{author}{Holtzman, B.K.}, \bibinfo{author}{King, D.S.},
  \bibinfo{author}{Kohlstedt, D.L.}, \bibinfo{year}{2012}.
\newblock \bibinfo{title}{Effects of stress-driven melt segregation on the
  viscosity of rocks}.
\newblock \bibinfo{journal}{Earth and Planetary Science Letters}
  \bibinfo{volume}{359-360}, \bibinfo{pages}{184--193}.
%Type = Article
\bibitem[{Holtzman and Kohlstedt(2007)}]{holtzman2007stressdriven}
\bibinfo{author}{Holtzman, B.K.}, \bibinfo{author}{Kohlstedt, D.L.},
  \bibinfo{year}{2007}.
\newblock \bibinfo{title}{Stress-driven melt segregation and strain
  partitioning in partially molten rocks: Effects of stress and strain}.
\newblock \bibinfo{journal}{Journal of Petrology} \bibinfo{volume}{48},
  \bibinfo{pages}{2379--2406}.
%Type = Article
\bibitem[{Humphreys and Dueker(1994)}]{humphreys1994western}
\bibinfo{author}{Humphreys, E.D.}, \bibinfo{author}{Dueker, K.},
  \bibinfo{year}{1994}.
\newblock \bibinfo{title}{Western {US} upper mantle structure}.
\newblock \bibinfo{journal}{J. Geophys. Res} \bibinfo{volume}{99}.
%Type = Article
\bibitem[{Humphreys et~al.(2003)Humphreys, Hessler, Dueker, Farmer, Erslev and
  Atwater}]{humphreys2003}
\bibinfo{author}{Humphreys, E.D.}, \bibinfo{author}{Hessler, E.},
  \bibinfo{author}{Dueker, K.}, \bibinfo{author}{Farmer, G.L.},
  \bibinfo{author}{Erslev, E.}, \bibinfo{author}{Atwater, T.},
  \bibinfo{year}{2003}.
\newblock \bibinfo{title}{How laramide-age hydration of north american
  lithosphere by the farallon slab controlled subsequent activity in the
  western united states}.
\newblock \bibinfo{journal}{International Geology Review} \bibinfo{volume}{45},
  \bibinfo{pages}{575--595}.
%Type = Article
\bibitem[{Katz et~al.(2006)Katz, Spiegelman and Holtzman}]{Katz:2006}
\bibinfo{author}{Katz, R.F.}, \bibinfo{author}{Spiegelman, M.},
  \bibinfo{author}{Holtzman, B.}, \bibinfo{year}{2006}.
\newblock \bibinfo{title}{The dynamics of melt and shear localization in
  partially molten aggregates}.
\newblock \bibinfo{journal}{Nature} \bibinfo{volume}{442},
  \bibinfo{pages}{676--679}.
%Type = Article
\bibitem[{Kelemen et~al.(1995)Kelemen, Whitehead, Aharonov and
  Jordahl}]{kelemen1995}
\bibinfo{author}{Kelemen, P.B.}, \bibinfo{author}{Whitehead, J.},
  \bibinfo{author}{Aharonov, E.}, \bibinfo{author}{Jordahl, K.},
  \bibinfo{year}{1995}.
\newblock \bibinfo{title}{Experiments on flow focusing in soluble porous media,
  with applications to melt extraction from the mantle}.
\newblock \bibinfo{journal}{Journal of Geophysical Research}
  \bibinfo{volume}{100}, \bibinfo{pages}{475--496}.
%Type = Article
\bibitem[{King et~al.(2011a)King, Hier-Majumder and Kohlstedt}]{King:2011p6140}
\bibinfo{author}{King, D.S.H.}, \bibinfo{author}{Hier-Majumder, S.},
  \bibinfo{author}{Kohlstedt, D.L.}, \bibinfo{year}{2011}a.
\newblock \bibinfo{title}{An experimental study of the effects of surface
  tension in homogenizing perturbations in melt fraction}.
\newblock \bibinfo{journal}{Earth Planet Sc Lett} , \bibinfo{pages}{1--13}.
%Type = Article
\bibitem[{King et~al.(2011b)King, Holtzman and Kohlstedt}]{King:2011p6141}
\bibinfo{author}{King, D.S.H.}, \bibinfo{author}{Holtzman, B.K.},
  \bibinfo{author}{Kohlstedt, D.L.}, \bibinfo{year}{2011}b.
\newblock \bibinfo{title}{An experimental investigation of the interactions
  between reaction-driven and stress-driven melt segregation. 1. application to
  mantle melt extraction}.
\newblock \bibinfo{journal}{Geochemistry, Geophysics, Geosystems}
  \bibinfo{volume}{submitted}, \bibinfo{pages}{1--38}.
%Type = Article
\bibitem[{Kohlstedt and Holtzman(2009)}]{Kohlstedt:2009}
\bibinfo{author}{Kohlstedt, D.L.}, \bibinfo{author}{Holtzman, B.K.},
  \bibinfo{year}{2009}.
\newblock \bibinfo{title}{Shearing melt out of the earth: An experimentalist's
  perspective on the influence of deformation on melt extraction}.
\newblock \bibinfo{journal}{Annual Reviews of Earth and Planetary Sciences}
  \bibinfo{volume}{37}, \bibinfo{pages}{561--93}.
%Type = Article
\bibitem[{Kreemer et~al.(2010)Kreemer, Blewitt and Bennett}]{Kreemer2010}
\bibinfo{author}{Kreemer, C.}, \bibinfo{author}{Blewitt, G.},
  \bibinfo{author}{Bennett, R.}, \bibinfo{year}{2010}.
\newblock \bibinfo{title}{Present-day motion and deformation of the colorado
  plateau}.
\newblock \bibinfo{journal}{Geophysical Research Letters} \bibinfo{volume}{37}.
%Type = Book
\bibitem[{Kusky et~al.(2007)Kusky, Windley and Zhai}]{kusky2007tectonic}
\bibinfo{author}{Kusky, T.M.}, \bibinfo{author}{Windley, B.F.},
  \bibinfo{author}{Zhai, M.G.}, \bibinfo{year}{2007}.
\newblock \bibinfo{title}{Tectonic evolution of the {North} {China} {Block}:
  from orogen to craton to orogen}.
\newblock \bibinfo{publisher}{Geological Society, London, Sepeial Publications
  280}.
%Type = Article
\bibitem[{Lee et~al.(2001)Lee, Yin, Rudnick and Jacobsen}]{lee2001}
\bibinfo{author}{Lee, C.T.}, \bibinfo{author}{Yin, Q.},
  \bibinfo{author}{Rudnick, R.L.}, \bibinfo{author}{Jacobsen, S.B.},
  \bibinfo{year}{2001}.
\newblock \bibinfo{title}{Preservation of ancient and fertile lithospheric
  mantle beneath the southwestern united states}.
\newblock \bibinfo{journal}{Nature} \bibinfo{volume}{411},
  \bibinfo{pages}{69--73}.
%Type = Article
\bibitem[{Lenardic et~al.(2003)Lenardic, Moresi and
  M{\"u}lhaus}]{lenardic2003longevity}
\bibinfo{author}{Lenardic, A.}, \bibinfo{author}{Moresi, L.N.},
  \bibinfo{author}{M{\"u}lhaus, H.}, \bibinfo{year}{2003}.
\newblock \bibinfo{title}{Longevity and stability of cratonic lithosphere:
  {Insights} from numerical simulations of coupled mantle convection and
  continental tectonics} .
%Type = Article
\bibitem[{Leroux et~al.(2008)Leroux, Tommasi and Vauchez}]{Leroux:2008p710}
\bibinfo{author}{Leroux, V.}, \bibinfo{author}{Tommasi, A.},
  \bibinfo{author}{Vauchez, A.}, \bibinfo{year}{2008}.
\newblock \bibinfo{title}{Feedback between melt percolation and deformation in
  an exhumed lithosphere--asthenosphere boundary}.
\newblock \bibinfo{journal}{Earth and Planetary Science Letters}
  \bibinfo{volume}{274}, \bibinfo{pages}{401--413}.
%Type = Article
\bibitem[{Levander and Miller(2012)}]{levandermiller12}
\bibinfo{author}{Levander, A.}, \bibinfo{author}{Miller, M.S.},
  \bibinfo{year}{2012}.
\newblock \bibinfo{title}{Evolutionary aspects of lithosphere discontinuity
  structure in the western u.s.}
\newblock \bibinfo{journal}{Geochemistry, Geophysics, Geosystems}
  \bibinfo{volume}{13}.
%Type = Article
\bibitem[{Levander et~al.(2011)Levander, Schmandt, Miller, Liu, Karlstrom,
  Crow, Lee and Humphreys}]{levander11}
\bibinfo{author}{Levander, A.}, \bibinfo{author}{Schmandt, B.},
  \bibinfo{author}{Miller, M.S.}, \bibinfo{author}{Liu, K.},
  \bibinfo{author}{Karlstrom, K.E.}, \bibinfo{author}{Crow, R.S.},
  \bibinfo{author}{Lee, C.T.A.}, \bibinfo{author}{Humphreys, E.D.},
  \bibinfo{year}{2011}.
\newblock \bibinfo{title}{Continuing colorado plateau uplift by
  delamination-style convective lithospheric downwelling}.
\newblock \bibinfo{journal}{Nature} \bibinfo{volume}{472},
  \bibinfo{pages}{461}.
%Type = Article
\bibitem[{Li et~al.(2007)Li, Yuan and Kind}]{Li_LAB_2007}
\bibinfo{author}{Li, X.}, \bibinfo{author}{Yuan, X.}, \bibinfo{author}{Kind,
  R.}, \bibinfo{year}{2007}.
\newblock \bibinfo{title}{The lithosphere-asthenosphere boundary beneath the
  western united states.}
\newblock \bibinfo{journal}{Geophysical Journal International}
  \bibinfo{volume}{170}.
%Type = Article
\bibitem[{Lin and Ritzwoller(2011)}]{lin_ritzwoller2011}
\bibinfo{author}{Lin, F.C.}, \bibinfo{author}{Ritzwoller, M.H.},
  \bibinfo{year}{2011}.
\newblock \bibinfo{title}{Helmholtz surface wave tomography for isotropic and
  azimuthally anisotropic structure}.
\newblock \bibinfo{journal}{{Geophysical} {Journal} {International}}
  \bibinfo{volume}{186}, \bibinfo{pages}{1104--1120}.
%Type = Article
\bibitem[{Lipman and Glazner(1991)}]{lipman1991}
\bibinfo{author}{Lipman, P.W.}, \bibinfo{author}{Glazner, A.F.},
  \bibinfo{year}{1991}.
\newblock \bibinfo{title}{Introduction to middle tertiary cordilleran volcanism
  -- magma sources and relations to regional tectonics}.
\newblock \bibinfo{journal}{J. Geophysical Research} \bibinfo{volume}{96},
  \bibinfo{pages}{193--13}.
%Type = Article
\bibitem[{Liu and Gurnis(2010)}]{liugurnis10}
\bibinfo{author}{Liu, L.}, \bibinfo{author}{Gurnis, M.}, \bibinfo{year}{2010}.
\newblock \bibinfo{title}{Dynamic subsidence and uplift of the colorado
  plateau}.
\newblock \bibinfo{journal}{Geology} \bibinfo{volume}{38},
  \bibinfo{pages}{663--666}.
%Type = Article
\bibitem[{Marchesi et~al.(2010)Marchesi, Griffin, Garrido, Bodinier, O'Reilly
  and Pearson}]{Marchesi2010}
\bibinfo{author}{Marchesi, C.}, \bibinfo{author}{Griffin, W.L.},
  \bibinfo{author}{Garrido, C.J.}, \bibinfo{author}{Bodinier, J.L.},
  \bibinfo{author}{O'Reilly, S.Y.}, \bibinfo{author}{Pearson, N.J.},
  \bibinfo{year}{2010}.
\newblock \bibinfo{title}{Persistence of mantle lithospheric re--os signature
  during asthenospherization of the subcontinental lithospheric mantle:
  insights from in situ isotopic analysis of sulfides persistence of mantle
  lithospheric re--os signature during asthenospherization of the
  subcontinental lithospheric mantle: insights from in situ isotopic analysis
  of sulfides from the ronda peridotite (southern spain)}.
\newblock \bibinfo{journal}{Contributions to Mineralogy and Petrology}
  \bibinfo{volume}{159}, \bibinfo{pages}{315--330}.
%Type = Article
\bibitem[{McKenzie(1984)}]{mckenzie1984}
\bibinfo{author}{McKenzie, D.}, \bibinfo{year}{1984}.
\newblock \bibinfo{title}{The generation and compaction of partially molten
  rock}.
\newblock \bibinfo{journal}{Journal of Petrology} \bibinfo{volume}{25},
  \bibinfo{pages}{713--765}.
%Type = Article
\bibitem[{Menzies et~al.(2007)Menzies, Xu, Zhang and Fan}]{menzies2007}
\bibinfo{author}{Menzies, M.}, \bibinfo{author}{Xu, Y.},
  \bibinfo{author}{Zhang, H.}, \bibinfo{author}{Fan, W.}, \bibinfo{year}{2007}.
\newblock \bibinfo{title}{Integration of geologygeophysics and geochemistry: A
  key to understanding the north china craton}.
\newblock \bibinfo{journal}{Lithos} \bibinfo{volume}{96},
  \bibinfo{pages}{1--21}.
%Type = Article
\bibitem[{Molnar and Jones(2004)}]{molnar2004test}
\bibinfo{author}{Molnar, P.}, \bibinfo{author}{Jones, C.},
  \bibinfo{year}{2004}.
\newblock \bibinfo{title}{A test of laboratory based rheological parameters of
  olivine from an analysis of late {Cenozoic} convective removal of mantle
  lithosphere beneath the {Sierra} {Nevada}, {California}, {USA}}.
\newblock \bibinfo{journal}{{Geophysical} {Journal} {International}}
  \bibinfo{volume}{156}, \bibinfo{pages}{555--564}.
%Type = Article
\bibitem[{Moucha et~al.(2008)Moucha, Forte, Rowley, Mitrovica, Simmons and
  Grand}]{moucha2008mantle}
\bibinfo{author}{Moucha, R.}, \bibinfo{author}{Forte, A.M.},
  \bibinfo{author}{Rowley, D.B.}, \bibinfo{author}{Mitrovica, J.X.},
  \bibinfo{author}{Simmons, N.A.}, \bibinfo{author}{Grand, S.P.},
  \bibinfo{year}{2008}.
\newblock \bibinfo{title}{Mantle convection and the recent evolution of the
  colorado plateau and the rio grande rift valley}.
\newblock \bibinfo{journal}{Geology} \bibinfo{volume}{36},
  \bibinfo{pages}{439--442}.
%Type = Article
\bibitem[{O'Reilly et~al.(2001)O'Reilly, Griffin, Djomani and
  Paul}]{reilly2001are}
\bibinfo{author}{O'Reilly, S.Y.}, \bibinfo{author}{Griffin, W.L.},
  \bibinfo{author}{Djomani, Y.H.M.}, \bibinfo{author}{Paul},
  \bibinfo{year}{2001}.
\newblock \bibinfo{title}{Are lithospheres forever? {Tracking} changes in the
  subcontinental lithospheric mantle through time}.
\newblock \bibinfo{journal}{GSA Today} \bibinfo{volume}{11},
  \bibinfo{pages}{4--10}.
%Type = Article
\bibitem[{Porreca and Selverstone(2006)}]{porreca2006pyroxenite}
\bibinfo{author}{Porreca, C.}, \bibinfo{author}{Selverstone, J.},
  \bibinfo{year}{2006}.
\newblock \bibinfo{title}{Pyroxenite xenoliths from the rio puerco volcanic
  field, new mexico: Melt metasomatism at the margin of the rio grande rift}.
\newblock \bibinfo{journal}{Geosphere} \bibinfo{volume}{2},
  \bibinfo{pages}{333--351;}.
%Type = Article
\bibitem[{Rau and Forsyth(2011)}]{RauForsyth11}
\bibinfo{author}{Rau, C.J.}, \bibinfo{author}{Forsyth, D.W.},
  \bibinfo{year}{2011}.
\newblock \bibinfo{title}{Melt in the mantle beneath the amagmatic zone,
  southern nevada}.
\newblock \bibinfo{journal}{Geology} \bibinfo{volume}{39},
  \bibinfo{pages}{975--978}.
%Type = Article
\bibitem[{Roy et~al.(2009)Roy, Jordan and Pederson}]{roy2009}
\bibinfo{author}{Roy, M.}, \bibinfo{author}{Jordan, T.H.},
  \bibinfo{author}{Pederson, J.}, \bibinfo{year}{2009}.
\newblock \bibinfo{title}{Cenozoic magmatism and rock uplift of the colorado
  plateau by warming of chemically buoyant lithosphere}.
\newblock \bibinfo{journal}{Nature} \bibinfo{volume}{459}.
%Type = Article
\bibitem[{Roy et~al.(2005)Roy, MacCarthy and Selverstone}]{roy2005}
\bibinfo{author}{Roy, M.}, \bibinfo{author}{MacCarthy, J.K.},
  \bibinfo{author}{Selverstone, J.}, \bibinfo{year}{2005}.
\newblock \bibinfo{title}{Upper mantle structure beneath the eastern colorado
  plateau and rio grande rift revealed by bouguer gravity, seismic velocities,
  and xenolith data}.
\newblock \bibinfo{journal}{Geochemistry Geophysics Geosystems}
  \bibinfo{volume}{6}.
%Type = Article
\bibitem[{Schmandt and Humphreys(2010)}]{schmandt2010}
\bibinfo{author}{Schmandt, B.}, \bibinfo{author}{Humphreys, E.D.},
  \bibinfo{year}{2010}.
\newblock \bibinfo{title}{Complex subduction and small-scale convection
  revealed by body-wave tomography of the western {United} {States} upper
  mantle}.
\newblock \bibinfo{journal}{Earth and Planetary Science Letters}
  \bibinfo{volume}{297}.
%Type = Article
\bibitem[{Schmeling and Wallner(2012)}]{schmeling12}
\bibinfo{author}{Schmeling, H.}, \bibinfo{author}{Wallner, H.},
  \bibinfo{year}{2012}.
\newblock \bibinfo{title}{Magmatic lithospheric heating and weakening during
  continental rifting: A simple scaling law, a 2-d thermomechanical rifting
  model and the east african rift system}.
\newblock \bibinfo{journal}{Geochemistry Geophysics Geosystems}
  \bibinfo{volume}{13}.
%Type = Article
\bibitem[{Sigoch(2011)}]{sigloch11}
\bibinfo{author}{Sigoch, K.}, \bibinfo{year}{2011}.
\newblock \bibinfo{title}{Mantle provinces under north america from
  multifrequency p wave tomography}.
\newblock \bibinfo{journal}{Geochemistry, Geophysics, Geosystems}
  \bibinfo{volume}{12}.
%Type = Article
\bibitem[{Silver and Holt(2002)}]{silver2002mantle}
\bibinfo{author}{Silver, P.G.}, \bibinfo{author}{Holt, W.E.},
  \bibinfo{year}{2002}.
\newblock \bibinfo{title}{The mantle flow field beneath western north america}.
\newblock \bibinfo{journal}{Science} \bibinfo{volume}{295},
  \bibinfo{pages}{1054--1057}.
%Type = Article
\bibitem[{Sine et~al.(2008)Sine, Wilson, Gao, Grand, Aster, Ni and
  Baldridge}]{sine2008mantle}
\bibinfo{author}{Sine, C.R.}, \bibinfo{author}{Wilson, D.},
  \bibinfo{author}{Gao, W.}, \bibinfo{author}{Grand, S.P.},
  \bibinfo{author}{Aster, R.}, \bibinfo{author}{Ni, J.},
  \bibinfo{author}{Baldridge, W.S.}, \bibinfo{year}{2008}.
\newblock \bibinfo{title}{Mantle structure beneath the western edge of the
  {Colorado} {Plateau}}.
\newblock \bibinfo{journal}{Geophysical Research Letters} \bibinfo{volume}{35}.
%Type = Article
\bibitem[{Smith(2000)}]{smith2000insights}
\bibinfo{author}{Smith, D.}, \bibinfo{year}{2000}.
\newblock \bibinfo{title}{Insights into the evolution of the uppermost
  continental mantle from xenolith localities on and near the {Colorado}
  {Plateau} and regional comparisons}.
\newblock \bibinfo{journal}{Journal of Geophysical Research}
  \bibinfo{volume}{105}, \bibinfo{pages}{16769--16781}.
%Type = Article
\bibitem[{Soustelle et~al.(2009)Soustelle, A.Tommasi, Bodinier, Garrido and
  A.Vauchez}]{soustelle2009}
\bibinfo{author}{Soustelle, V.}, \bibinfo{author}{A.Tommasi},
  \bibinfo{author}{Bodinier, J.L.}, \bibinfo{author}{Garrido, C.J.},
  \bibinfo{author}{A.Vauchez}, \bibinfo{year}{2009}.
\newblock \bibinfo{title}{Deformation and reactive melttransport in the mantle
  lithosphere above a large-scale partial melting domain: the ronda peridotite
  massif, southern spain}.
\newblock \bibinfo{journal}{Journal of Petrology} \bibinfo{volume}{50},
  \bibinfo{pages}{1235--1266}.
%Type = Article
\bibitem[{Spiegelman(1993)}]{spieg93}
\bibinfo{author}{Spiegelman, M.}, \bibinfo{year}{1993}.
\newblock \bibinfo{title}{Flow in deformable porous media. part 1 simple
  analysis}.
\newblock \bibinfo{journal}{Journal of Fluid Mechanics} \bibinfo{volume}{247},
  \bibinfo{pages}{17--38}.
%Type = Article
\bibitem[{Spiegelman(2003)}]{spieg2003}
\bibinfo{author}{Spiegelman, M.}, \bibinfo{year}{2003}.
\newblock \bibinfo{title}{Linear analysis of melt band formation by simple
  shear}.
\newblock \bibinfo{journal}{Geochemistry Geophysics Geosystems}
  \bibinfo{volume}{4}.
%Type = Article
\bibitem[{Swanberg and Morgan(1985)}]{swanberg_morgan}
\bibinfo{author}{Swanberg, C.A.}, \bibinfo{author}{Morgan, P.M.P.},
  \bibinfo{year}{1985}.
\newblock \bibinfo{title}{Silica heat flow estimates and heat flow in the
  colorado plateau and adjacent areas}.
\newblock \bibinfo{journal}{Journal of Geodynamics} \bibinfo{volume}{3},
  \bibinfo{pages}{65--85}.
%Type = Article
\bibitem[{Takei and Hier-Majumder(2009)}]{Takei:2009p2204}
\bibinfo{author}{Takei, Y.}, \bibinfo{author}{Hier-Majumder, S.},
  \bibinfo{year}{2009}.
\newblock \bibinfo{title}{A generalized formulation of interfacial tension
  driven fluid migration with dissolution/precipitation}.
\newblock \bibinfo{journal}{Earth and Planetary Science Letters}
  \bibinfo{volume}{288}, \bibinfo{pages}{138--148}.
%Type = Article
\bibitem[{Takei and Holtzman(2009)}]{Takei:2009p1421}
\bibinfo{author}{Takei, Y.}, \bibinfo{author}{Holtzman, B.K.},
  \bibinfo{year}{2009}.
\newblock \bibinfo{title}{Viscous constitutive relations of solid-liquid
  composites in terms of grain boundary contiguity: 3. causes and consequences
  of viscous anisotropy}.
\newblock \bibinfo{journal}{J. Geophys. Res.} \bibinfo{volume}{114},
  \bibinfo{pages}{23}.
%Type = Article
\bibitem[{Tian et~al.(2009)Tian, Sigloch and Nolet}]{tian09}
\bibinfo{author}{Tian, Y.}, \bibinfo{author}{Sigloch, K.},
  \bibinfo{author}{Nolet, G.}, \bibinfo{year}{2009}.
\newblock \bibinfo{title}{Multiple-frequency sh-wave tomography of the western
  us upper mantle}.
\newblock \bibinfo{journal}{Geophysical Journal International}
  \bibinfo{volume}{178}, \bibinfo{pages}{1384--1402}.
%Type = Article
\bibitem[{Wagner et~al.(2010)Wagner, Forsyth, Fouch and James}]{wagner10}
\bibinfo{author}{Wagner, L.}, \bibinfo{author}{Forsyth, D.W.},
  \bibinfo{author}{Fouch, M.J.}, \bibinfo{author}{James, D.E.},
  \bibinfo{year}{2010}.
\newblock \bibinfo{title}{Detailed three-dimensional shear wave velocity
  structure of the northwestern united states from rayleigh wave tomography}.
\newblock \bibinfo{journal}{Earth and Planetary Science Letters}
  \bibinfo{volume}{299}.
%Type = Unpublished
\bibitem[{Wang et~al.()Wang, Forsyth, Rau, Carriero, Schmandt, Gaherty and
  Savage}]{Forsyth12Isabella}
\bibinfo{author}{Wang, Y.}, \bibinfo{author}{Forsyth, D.},
  \bibinfo{author}{Rau, C.}, \bibinfo{author}{Carriero, N.},
  \bibinfo{author}{Schmandt, B.}, \bibinfo{author}{Gaherty, J.},
  \bibinfo{author}{Savage, B.}, .
\newblock \bibinfo{title}{Fossil slabs attached to unsubducted fragments of the
  farallon plate}.
\newblock \bibinfo{note}{Submitted for publication}.
%Type = Article
\bibitem[{Wannamaker et~al.(2008)Wannamaker, Hasterok, Johnston, Stodt, Hall,
  Sodergren, Pellerin, Maris, Doerner and Unsworth}]{wannamaker08}
\bibinfo{author}{Wannamaker, P.E.}, \bibinfo{author}{Hasterok, D.P.},
  \bibinfo{author}{Johnston, J.M.}, \bibinfo{author}{Stodt, J.A.},
  \bibinfo{author}{Hall, D.B.}, \bibinfo{author}{Sodergren, T.L.},
  \bibinfo{author}{Pellerin, L.}, \bibinfo{author}{Maris, V.},
  \bibinfo{author}{Doerner, W.M.}, \bibinfo{author}{Unsworth, K.A.G.M.J.},
  \bibinfo{year}{2008}.
\newblock \bibinfo{title}{Lithospheric dismemberment and magmatic processes of
  the great basin--colorado plateau transition, utah, implied from
  magnetotellurics}.
\newblock \bibinfo{journal}{Geochemistry Geophysics Geosystems}
  \bibinfo{volume}{9}.
%Type = Article
\bibitem[{Wenrich et~al.(1995)}]{wenrich1995spatial}
\bibinfo{author}{Wenrich, K.J.}, et~al., \bibinfo{year}{1995}.
\newblock \bibinfo{title}{Spatial migration and compositional changes of
  {Miocene}-{Quaternary} magmatism in the western {Grand} {Canyon}}.
\newblock \bibinfo{journal}{Journal of Geophysical Research}
  \bibinfo{volume}{100}, \bibinfo{pages}{10417--10440}.
%Type = Article
\bibitem[{West et~al.(2009)West, Fouch, Roth and
  Elkins-Tanton}]{west2009vertical}
\bibinfo{author}{West, J.D.}, \bibinfo{author}{Fouch, M.J.},
  \bibinfo{author}{Roth, J.B.}, \bibinfo{author}{Elkins-Tanton, L.T.},
  \bibinfo{year}{2009}.
\newblock \bibinfo{title}{Vertical mantle flow associated with a lithospheric
  drop beneath the {Great} {Basin}, {Nature} {Geoscience}}.
\newblock \bibinfo{journal}{DOI:10} \bibinfo{volume}{1038/}.
%Type = Article
\bibitem[{West et~al.(2004)West, Ni, Baldridge, Wilson, Aster, Gao and
  Grand}]{west2004crust}
\bibinfo{author}{West, M.}, \bibinfo{author}{Ni, J.},
  \bibinfo{author}{Baldridge, W.}, \bibinfo{author}{Wilson, D.},
  \bibinfo{author}{Aster, R.}, \bibinfo{author}{Gao, W.},
  \bibinfo{author}{Grand, S.}, \bibinfo{year}{2004}.
\newblock \bibinfo{title}{Crust and upper mantle shear-wave structure of the
  southwest {United} {States}: {Implications} for rifting and support for high
  elevation}.
\newblock \bibinfo{journal}{J. Geophys. Res.} \bibinfo{volume}{109}.
%Type = Article
\bibitem[{van Wijk et~al.(2010)van Wijk, Baldridge, van Hunen, Goes, Aster,
  Coblentz, Grand and Ni}]{vanwijk10}
\bibinfo{author}{van Wijk, J.}, \bibinfo{author}{Baldridge, W.},
  \bibinfo{author}{van Hunen, J.}, \bibinfo{author}{Goes, S.},
  \bibinfo{author}{Aster, R.}, \bibinfo{author}{Coblentz, D.},
  \bibinfo{author}{Grand, S.}, \bibinfo{author}{Ni, J.}, \bibinfo{year}{2010}.
\newblock \bibinfo{title}{Small-scale convection at the edge of the colorado
  plateau: Implications for topography, magmatism, and evolution of proterozoic
  lithosphere}.
\newblock \bibinfo{journal}{Geology} \bibinfo{volume}{38},
  \bibinfo{pages}{611--614}.
%Type = Article
\bibitem[{W{\"o}lbern et~al.(2012)W{\"o}lbern, R{\"u}mpker, Link and
  Sodoudi}]{wolbern12}
\bibinfo{author}{W{\"o}lbern, I.}, \bibinfo{author}{R{\"u}mpker, G.},
  \bibinfo{author}{Link, K.}, \bibinfo{author}{Sodoudi, F.},
  \bibinfo{year}{2012}.
\newblock \bibinfo{title}{Melt infiltration of the lower lithosphere beneath
  the tanzania craton and the albertine rift inferred from s receiver
  functions}.
\newblock \bibinfo{journal}{GEOCHEMISTRY GEOPHYSICS GEOSYSTEMS, in press} .
%Type = Article
\bibitem[{Xue and Allen(2010)}]{xueallen2010}
\bibinfo{author}{Xue, M.}, \bibinfo{author}{Allen, R.M.}, \bibinfo{year}{2010}.
\newblock \bibinfo{title}{Mantle structure beneath the western united states
  and its implications for convection processes}.
\newblock \bibinfo{journal}{Journal of Geophysical Research}
  \bibinfo{volume}{B07303}.
%Type = Article
\bibitem[{Yang et~al.(2008)Yang, Ritzwoller, Lin, Moschetti and
  Shapiro}]{Yang2008}
\bibinfo{author}{Yang, Y.}, \bibinfo{author}{Ritzwoller, M.H.},
  \bibinfo{author}{Lin, F.C.}, \bibinfo{author}{Moschetti, M.P.},
  \bibinfo{author}{Shapiro, N.M.}, \bibinfo{year}{2008}.
\newblock \bibinfo{title}{Structure of the crust and uppermost mantle beneath
  the western united states revealed by ambient noise and earthquake
  tomography}.
\newblock \bibinfo{journal}{Journal of Geophysical Research}
  \bibinfo{volume}{113}.
%Type = Article
\bibitem[{Zandt and Carrigan(1993)}]{ZandtCarrigan93}
\bibinfo{author}{Zandt, G.}, \bibinfo{author}{Carrigan, C.},
  \bibinfo{year}{1993}.
\newblock \bibinfo{title}{Small-scale convective instability and upper mantle
  viscosity under california small-scale convective instability and upper
  mantle viscosity under california small-scale convective instability and
  upper mantle viscosity under california small-scale convective instability
  and upper mantle viscosity under california}.
\newblock \bibinfo{journal}{Science} \bibinfo{volume}{261},
  \bibinfo{pages}{460--463}.
%Type = Article
\bibitem[{Zhong et~al.(2000)Zhong, Zuber, Moresi and Gumis}]{zhong2000}
\bibinfo{author}{Zhong, S.}, \bibinfo{author}{Zuber, M.T.},
  \bibinfo{author}{Moresi, L.}, \bibinfo{author}{Gumis, M.},
  \bibinfo{year}{2000}.
\newblock \bibinfo{title}{Role of temperature-dependent viscosity and surface
  plates in spherical shell models of mantle convection}.
\newblock \bibinfo{journal}{Journal of Geophysical Research}
  \bibinfo{volume}{105}, \bibinfo{pages}{11,063--11,082}.

\end{thebibliography}
%\bibliographystyle{agufull}
%\bibliographystyle{model2-names}

%% ------------------------------------------------------------------------ %%
%
%  END ARTICLE
%
%% ------------------------------------------------------------------------ %%

%% Enter Figures and Tables here:

\pagestyle{empty}

\begin{figure}[!hhh] 
\begin{center}
\includegraphics[width=4.5in]{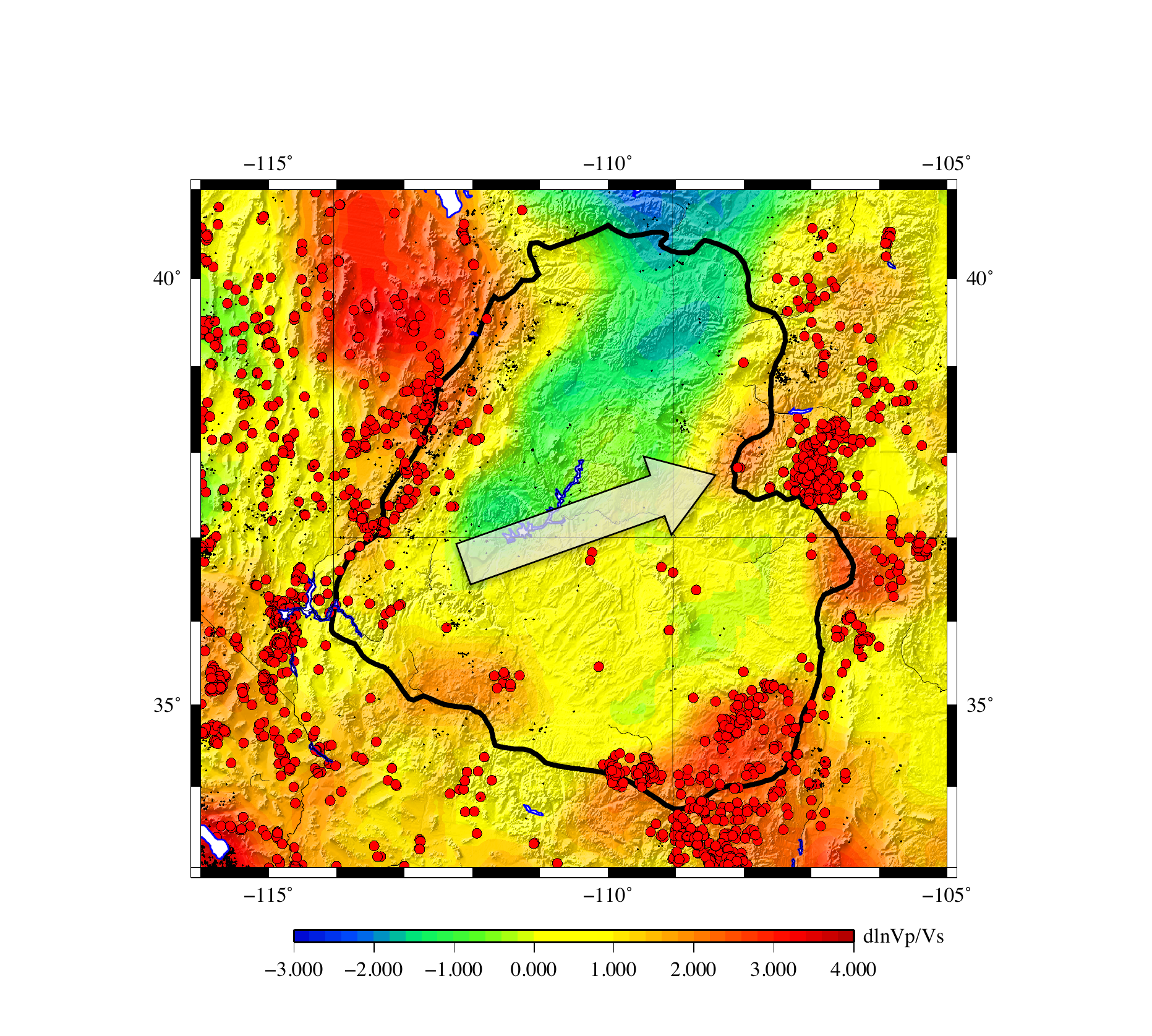}
\caption{ A composite image of the Colorado Plateau region (thick black line marks the physiographic boundary of the plateau) with published $V_{p}/V_{s}$ ratios (color image) at ~90 km depth, modified from \citet{schmandt2010}. Large red circles indicate Cenozoic volcanic rocks in the NAVDAT database and small black dots indicate seismicity.  Greater seismicity and magmatism at the margins of the plateau coincides with regions of higher $V_{p}/V_{s}$ ratio relative to the interior. Large arrow indicates the direction of motion of the underlying asthenosphere in a reference frame stationary with respect to North America inferred from a combination of kinematic data and seismic anisotropy \citep{silver2002mantle}. }
\end{center}
\end{figure}
\begin{figure}
\begin{center}
\includegraphics[width=6 in]{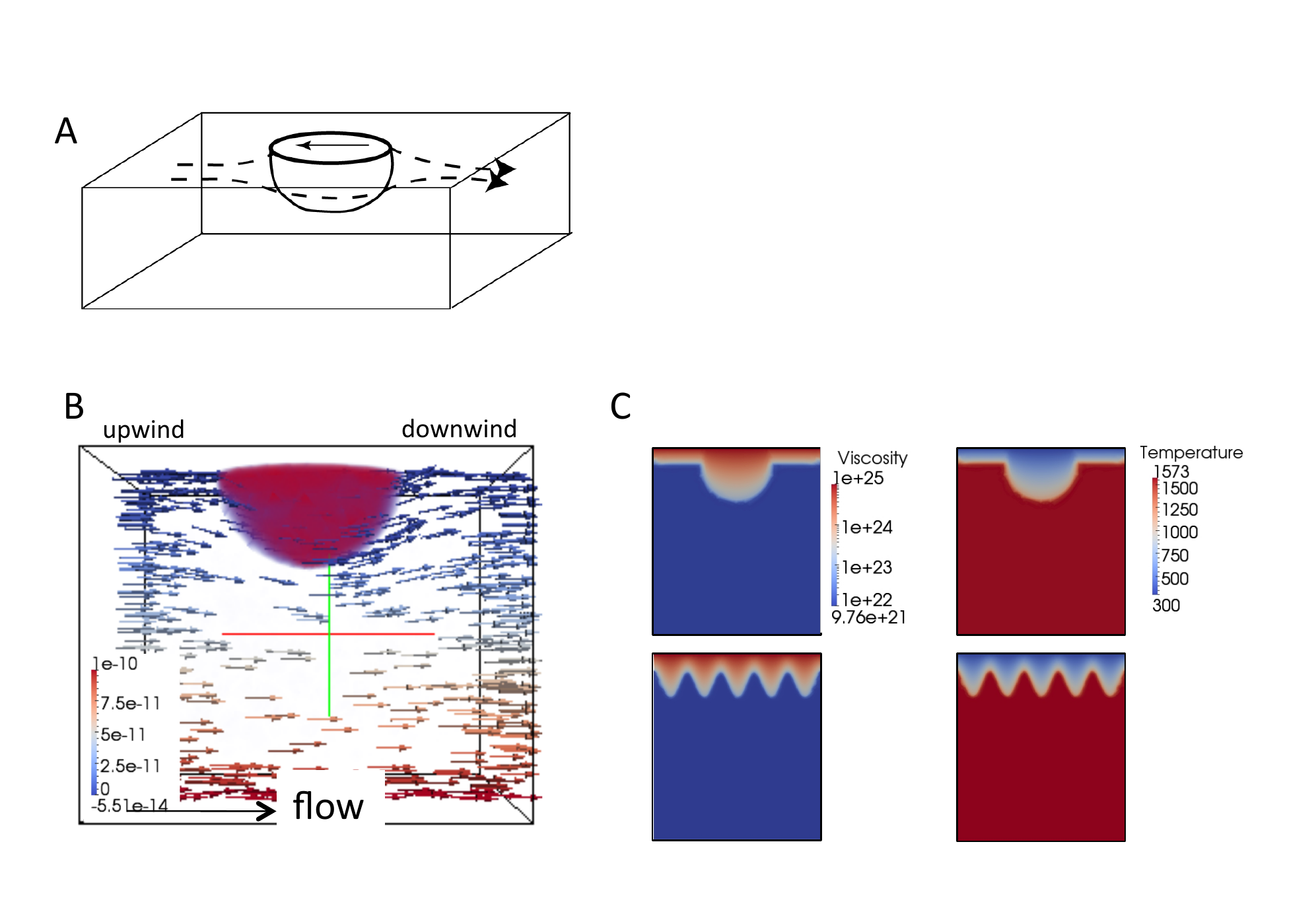}
\caption{A. Cartoon of model showing a hemispherical region of thick, high-viscosity lithosphere protruding into lower-viscosity asthenosphere that flows from left to right. B. Kinematic boundary conditions at the base of the model drive flow in the asthenosphere around the high-viscosity protrusion, represented as a red hemispherical volume.  The model domain is 1200 km by 1200 km by 1000 km (depth), with the hemisphere radius=300 km in this calculation. The bottom boundary is driven from left to right at $\sim10^{-9}$ m/s (red arrows) while the top boundary is held fixed.  The steady-state flow field is shown for a slice through the center of the model, with velocity vectors color-coded by magnitude in m/s.  C. Initial viscosity (Pa s) and temperature (K) structures in 2D models for a semicircular lithospheric protrusion (top) and for periodic ridges at the base of the plate (bottom).}
\end{center}
\end{figure}
\begin{figure}
\begin{center}
\includegraphics[width=6 in]{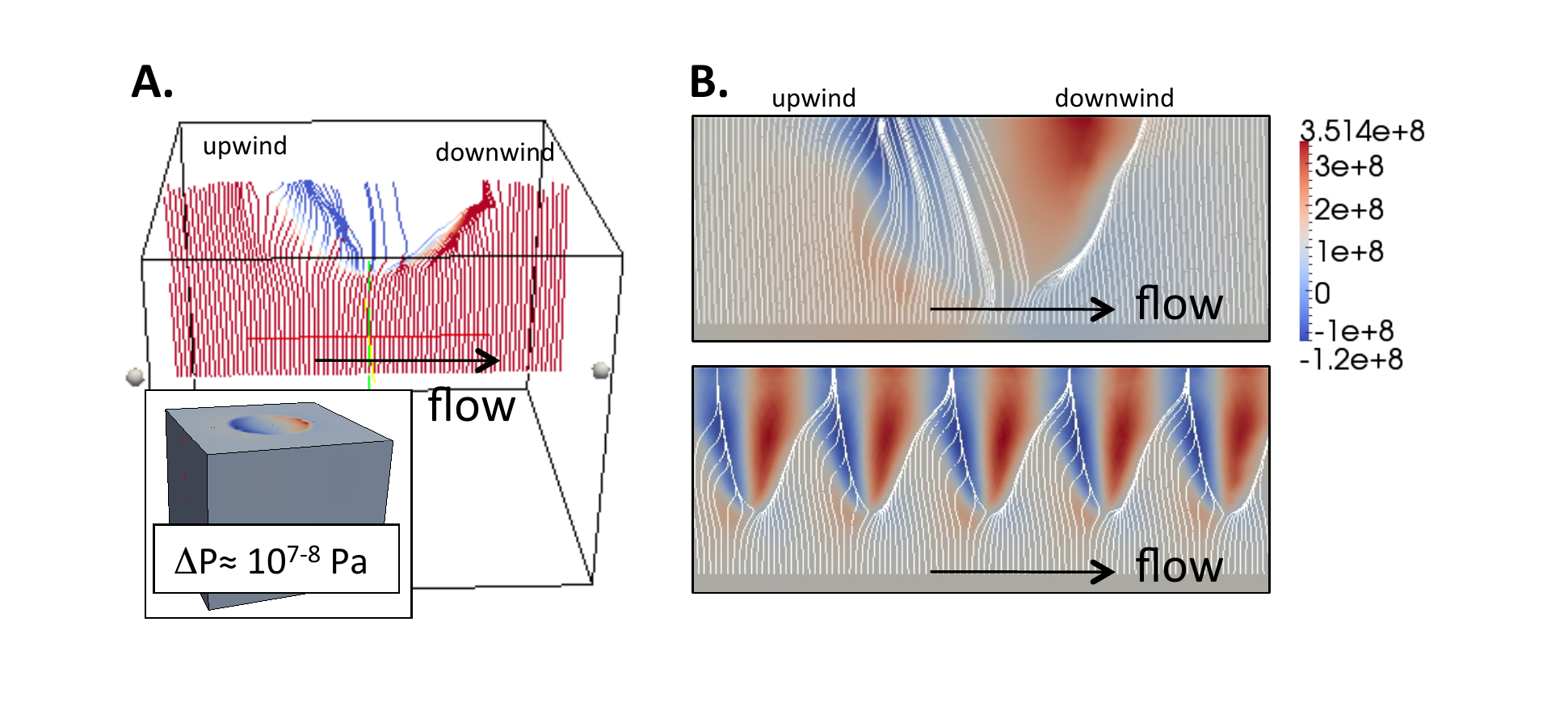}
\caption{A. Melt extraction streamlines calculated from the flow field in Figure 2B, colored red when the line is outside the hemispherical keel and blue when it is within the keel.  The pattern of melt-extraction illustrates primarily upward flow, dominated by buoyancy of the melt, but with significant lateral diversions due to dynamic pressure gradients (illustrated in the inset) that drive melt into the keel on the ``upwind" side and suck melt away from the keel on its ``downwind" side. B. Melt extraction streamlines calculated for the semicircle and periodic ridges in Figure 2C, showing the persistent asymmetry in dynamic pressures (in Pa) with calculated melt streamlines superimposed in white.}
\end{center}
\end{figure}

\begin{figure}
\begin{center}
\noindent\includegraphics[width=6 in]{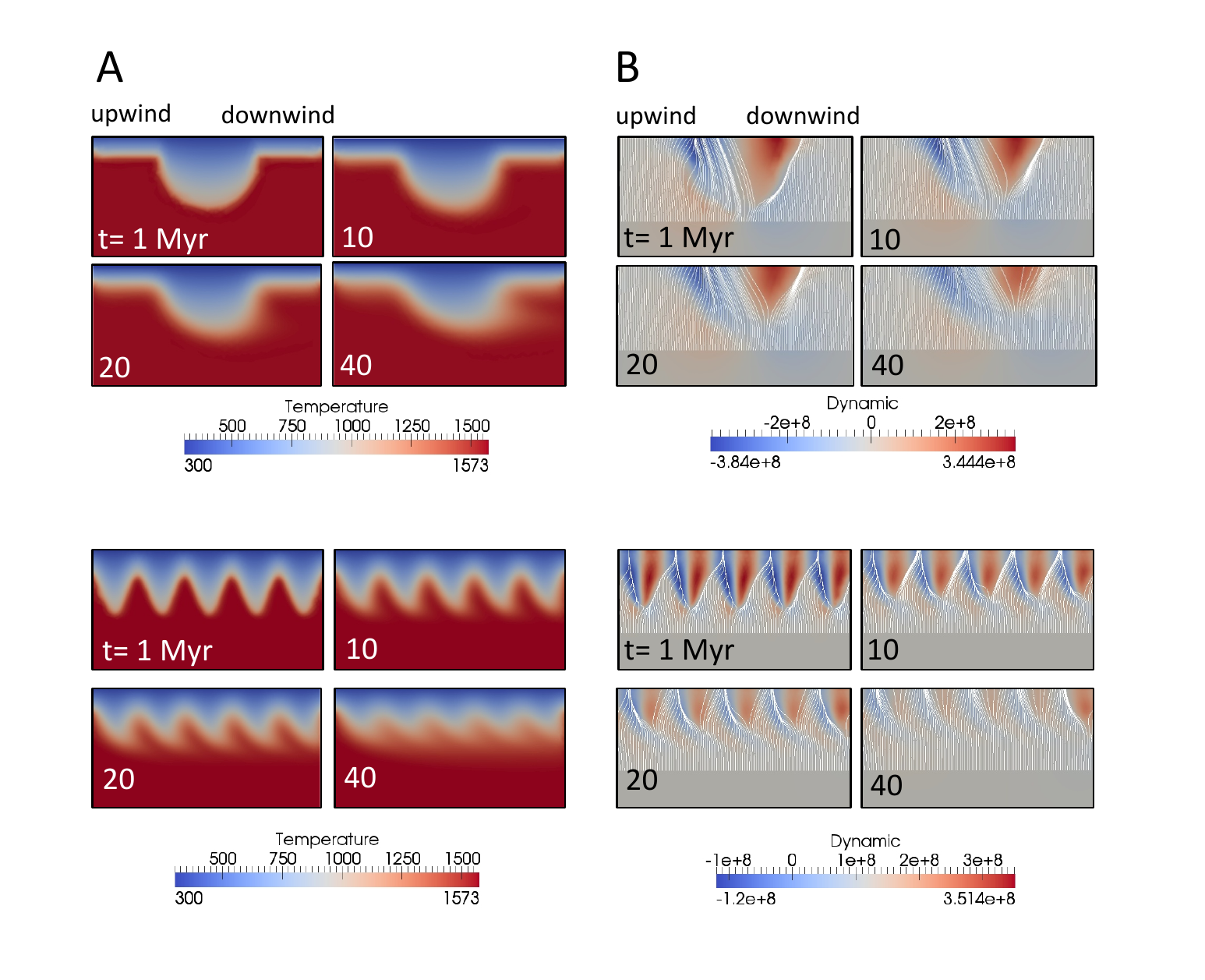}
\caption{A. Temporal evolution of the temperature (A, in K) and dynamic pressure (B, in Pa) fields and accompanying changes in melt extraction through time.  Results are shown for semicircular protrusions (top panels) and periodic ridges (bottom panels). The effects of temperature dependent viscosity and a combination of advective and conductive heating may be seen for tens of millions of years, leading to a persistent upwind-downwind asymmetry in both dynamic pressures and in melt extraction.}
\end{center}
\end{figure}

\begin{figure}
\begin{center}
\includegraphics[width=5.5 in]{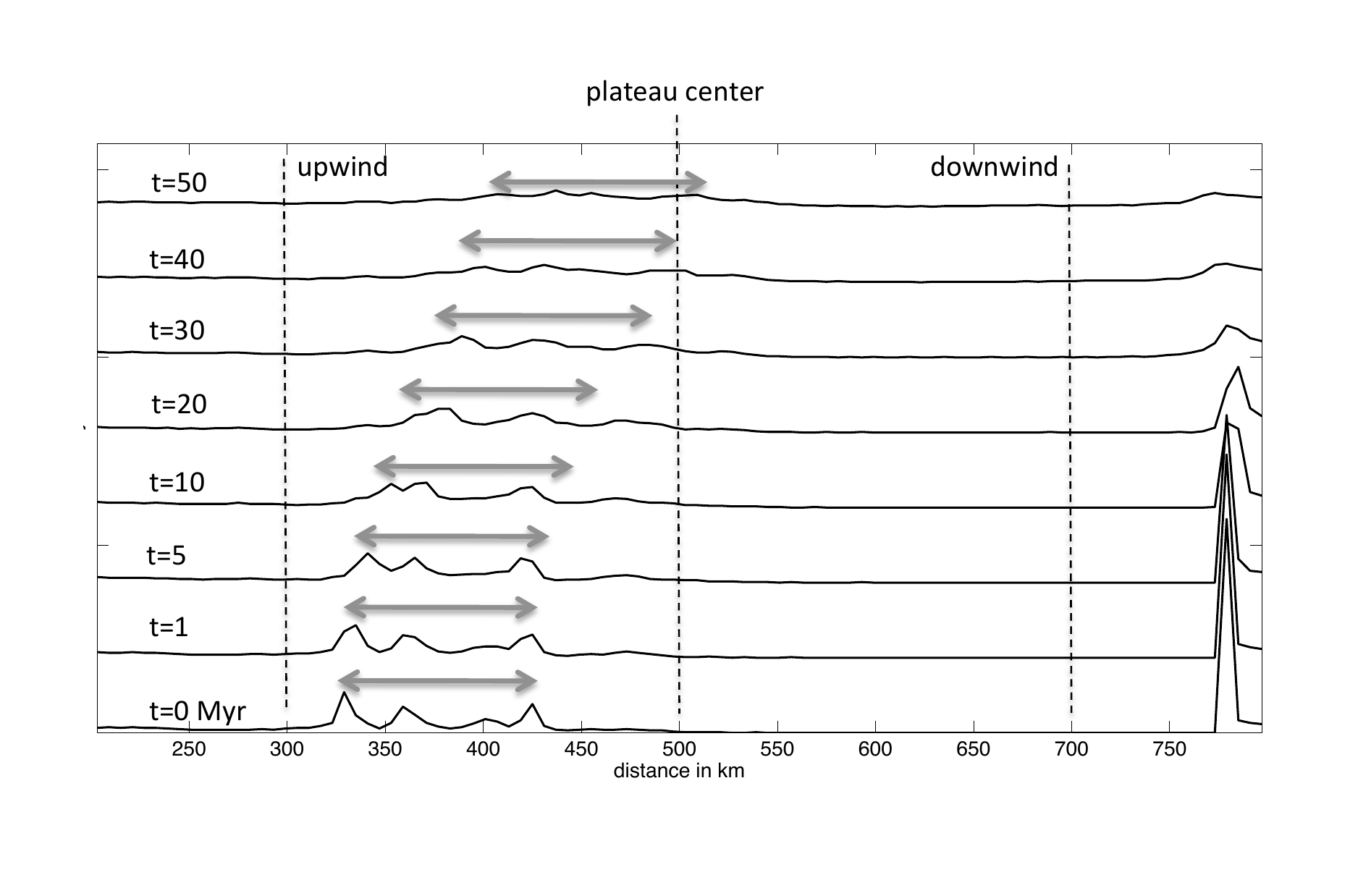}
\caption{Plot of normalized streamline density at the surface (a measure of melt-focusing defined as the number of streamlines per km at the surface, normalized by the value at depth below the keel) as a function of time for the semicircular protrusion model shown in Figure 4B.  The vertical dashed lines indicate the edges (at 300 and 700 km) and the center (at 500 km) of the semi-circular protrusion.  This calculation shows the inward migration of the zone of melt-focusing on the upwind side of the protrusion (indicated by grey double arrows) toward the center of the protrusion.  On the downwind side, there is a persistent exclusion of melt streamlines (and lack of migration).}
\end{center}
\end{figure}

\begin{figure}
\begin{center}
\includegraphics[width=6 in]{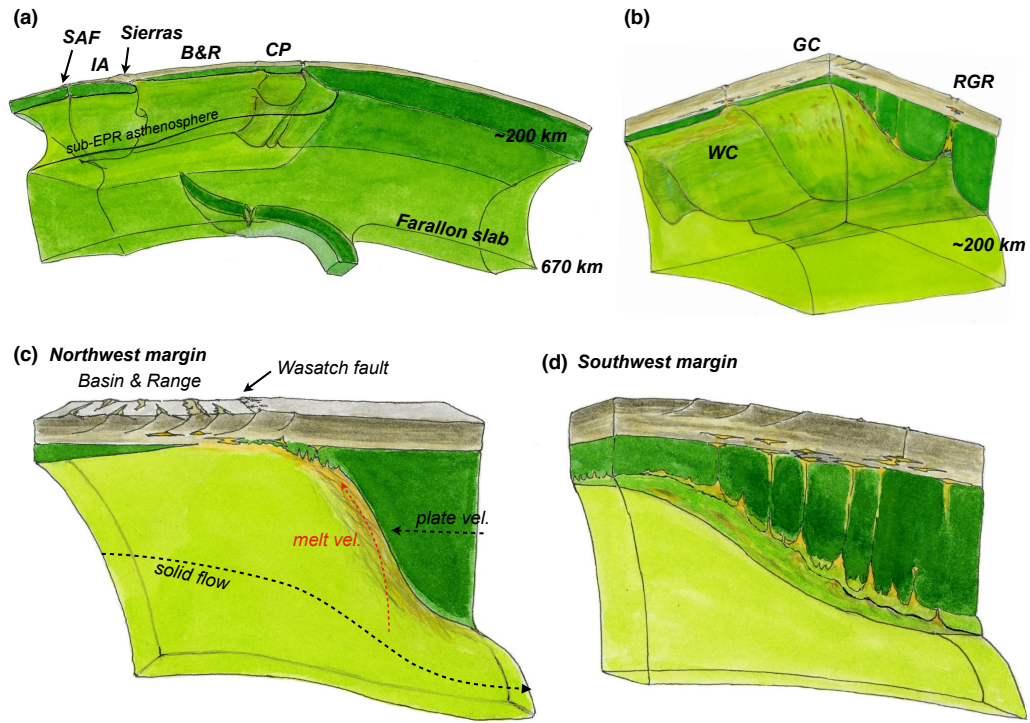}
\caption{(a) Cartoon illustrating the geodynamic context for the Colorado
Plateau within the western US.  Cenozoic evolution of the western US is characterized by removal and fragmentation of the Farallon slab, followed by possible replacement of the sub-lithospheric mantle by warm, sub-EPR asthenosphere \citep[e.g., ][]{moucha2008mantle}; WC=Wyoming craton; CP=Colorado Plateau; SAF=San Andreas Fault; B\&R=Basin and Range; IA=Isabella Anomaly, a possible remnant of the Farallon slab \citep{Forsyth12Isabella}.  (b) The keel of the Colorado Plateau viewed from below, looking roughly NE (GC=The Grand Canyon; RGR=Rio Grande Rift). The keel may have significantly different character on the LAB at the NW margin as compared to other margins, as shown in (c) and (d). (c) We hypothesize the presence of stress-driven melt segregation (SDS) along a steep LAB at the high strain-rate NW margin, characterized by sharp lateral contrasts in seismic wave speed and electrical conductivity.  (d) At lower strain-rate margins, such as the SW margin, we hypothesize more diffuse zones of modification/corrosion; the absence of large degrees of SDS corresponds to the behavior modeled in this paper, with melt migrating dominantly upward but modified by dynamic pressure gradients in the keel. }
\end{center}
\end{figure}

% ---------------

\newpage
\begin{table}
\begin{center}
\begin{tabular}{c c l}
\hline
 Symbol  & Variable & Values used\\
\hline
  $\phi$  & porosity & $< 1\%$ (low porosity limit)\\
  $\kappa$  & solid permeability  & $10^{-15}$ to $10^{-12}$ m$^2$\\
  $\rho_{f}$  & melt density & 2800 kg/m$^3$ \\ 
  $\rho_{s}$  & solid density & 3300 kg/m$^3$   \\
  $\mathbf{u}$  & melt velocity  & calculated in Eqn. 1 \\
  $\mathbf{v}$  & solid velocity  & $0$ to $1.6^{9}$ m/s   \\
  $\eta$, $\eta_{l}$ $\eta_{a}$ & solid shear viscosity &  $10^{18}$ to $10^{25}$ Pa s \\
  $\zeta$ & solid bulk viscosity &  $0$; no compaction in melt-streamline calculations \\
  $\mu$  & melt viscosity & 1 Pa s   \\
  $P$  & fluid pressure &   hydrostatic pressures up to $4\times10^{10} Pa$ \\
  $\mathbf{g}$ & acceleration due to gravity & $9.82$ m/s$^2$ \\
  $\eta_{0}$ & reference viscosity & $10^{22}$ Pa s\\
  $E$  & activation parameter $E/R(\Delta T)^{2}$ & $5.4 \times 10^{-3}$ /K \\
\hline
\end{tabular}
\caption{Physical parameters in models}
\end{center}
\end{table}

\end{document}